\pdfoutput=1
\documentclass[runningheads]{llncs}

\usepackage[english]{babel}

\usepackage{cite}

\usepackage[english]{babel}

\usepackage{amsmath}
\usepackage{amssymb}
\usepackage{graphicx}
\usepackage{float} 
\usepackage[colorlinks=true, allcolors=blue]{hyperref}

\usepackage{algorithm}
\usepackage{algorithmicx}
\usepackage{algpseudocode}
\usepackage{listings}
\usepackage{inconsolata} 

\usepackage{xcolor}

\usepackage[final]{changes}
 
\definechangesauthor[name=copy, color=purple]{copy}
\definechangesauthor[name=Review1, color=blue]{R1}
\definechangesauthor[name=Review2, color=red]{R2}
\definechangesauthor[name=Review3, color=orange]{R3}
\definechangesauthor[name=Ours, color=green]{Ours}

\newtheorem{concept}{Concept}

\newtheorem{observation}{Observation}

\title{Annotation-guided AoS-to-SoA conversions and GPU offloading with data views in C++
 \thanks{
  This is an extended version of the 2024 PPAM paper \emph{Compiler support for semi-manual AoS-to-SoA conversions with data views}.
 }
}
\titlerunning{Annotation-guided AoS-to-SoA conversions}

\author{Pawel K.~Radtke\inst{1}\orcidID{0009-0009-8613-3632} \and Tobias
Weinzierl\inst{1}\orcidID{0000-0002-6208-1841}}

\authorrunning{P. Radtke and T. Weinzierl}

\institute{Department of Computer Science, Durham University, Durham, United
Kingdom \email{\{pawel.k.radtke,tobias.weinzierl\}@durham.ac.uk}\\
\url{https://scicomp.webspace.durham.ac.uk}}

\begin{document}
\maketitle

\begin{abstract}
  The C++ programming language provides classes and structs as fundamental modeling entities. 
Consequently, C++ code tends to favour array-of-structs (AoS) for encoding data sequences, even though
structure-of-arrays (SoA) yields better performance for some calculations. 
We propose a C++ language extension based on attributes that allows developers to guide the compiler in selecting memory arrangements, i.e.~to select the optimal choice between AoS and SoA dynamically depending on both the execution context and algorithm step.
The compiler can then automatically convert data into the preferred format prior to the calculations and convert results back afterward. 
The compiler handles all the complexity of determining which data to convert and how to manage data transformations.
Our implementation realises the compiler-extension for the new annotations in Clang and demonstrates their effectiveness through a smoothed particle hydrodynamics (SPH) code, which we evaluate on an Intel CPU, an ARM CPU, and a Grace-Hopper GPU. 
While the separation of concerns between data structure and operators is elegant and provides performance improvements, the new annotations do not eliminate the need for performance engineering. 
Instead, they challenge conventional performance wisdom and necessitate rethinking approaches how to write efficient implementations.

\end{abstract}

\begin{keywords}
  Array-of-structs, struct-of-arrays, memory layout transformations, data views,
  compiler, vectorisation, GPU offloading
\end{keywords}

\section{Introduction}
\label{section:introduction}

%
%
Data movements dominate the runtime on supercomputers.
On a state-of-the-art compute node, moving data through the cache hierarchy often lasts significantly longer than the actual arithmetics, while accelerator offloading notoriously suffers from data transfer overheads.
Data movement penalties are amplified by irregular and scattered memory access.
Despite efforts in modern algorithmics to mitigate this problem (e.g.~by switching to higher-order schemes), many codes remain memory-bound.
Computational paradigms such as multiscale modelling and adaptive meshing amplify this effect.
Their core ideas do not align well with the trends in memory architectures. 
There is no sign that the memory gap between data movement and compute speed closes any time soon.
It is widening~\cite{Dongarra:2011:ExascaleSoftwareRoadmap}.

%
%
\added[id=copy]{Modern high-level languages} such as C/C++ \deleted[id=Ours]{run} risk exacerbating the arising problems:
They \added[id=copy]{provide us with the concept of structures (structs) to model data such as particles, mesh cells, or entries of a multiphysics partial differential equations. 
Structs are the natural modelling vocabulary for many algorithms.
Due to \texttt{struct}, the C++ language leans towards an array-of-structs (AoS) storage for sequences of objects, aka instances of structs.
The struct---or class in object-oriented terminology---is the primary modelling entity for the programmer}~\cite{Hirzel:2007:DataLayoutForOO,Homann:2018:SoAx,Reinders:2016:XeonPhi,Jubertie:2018:DataLayoutAbstractionLayers}.
In many algorithms, structs are not only the natural modelling langugae, they are also convenient to organise (partial) data exchange accross process boundaries through message passing, sorting, or to deliver maintainable and extendable code bases.
Unfortunately, thinking and programming in structs tends to add further data scattering---if not all elements of a struct are required within a compute-heavy subprogram (computational kernel), each cache line load brings data into the cache that is not required at this point, i.e.~the cache's capability to hold relevant data is reduced---and non-coalesced memory accesses which in turn challenge vectorisation.

%
%
As developers can not alter the hardware and typically do not want to rewrite AoS code, they can resort to a \emph{temporary data reorganisation in memory}:
Data are converted back and forth between AoS and structures of arrays (SoA), depending on the compute task ahead.
This is a counter-intuitive strategy:
We introduce further memory movements in a regime already constrained by them.
\added[id=copy]{Yet, as implementations over SoA often outperform their AoS counterparts---once they enable efficient usage of vector instructions~\cite{Gallard:2020:ExaHyPEVectorisation,Intel:MemoryLayoutTransformations,Springer:2018:SoALayout}, manage to exploit GPUs \cite{Che:2011:Dymaxion,Majeti:2014:CompilerDrivenDataLayoutTransformations} and become less sensitive to cache effects than  AoS~\cite{Homann:2018:SoAx,Hundt:2006:StructureLayoutOptimisation,Springer:2018:SoALayout}---}many codes have a sweet spot, e.g.~in terms of problem sizes, where a reordering pays off.
To bring the sweetspot forward, we introduce data \emph{views} (see preparatory work in \cite{Radtke:2024:AoS2SoA}): 
Any temporary code data reordering includes exclusively those struct members that are read from or written to.
Other struct members remain untouched. 
Logically, we work with a view over the data type that exposes only specific members.

%
%
To streamline programming following this implementation idiom, four properties are desirable: 
First, all transformations must be minimally invasive, avoiding major code changes that ripple through the codebase.
Second, they should leave core routines with domain knowledge (e.g., physical models) written in AoS untouched.
Third, data conversions should be hidden from the programmer and the code hence remain usable in both AoS and SoA contexts.
Finally, the result should be optimal with minimal data movements.
We therefore generalise the notion of a view, distinguishing read from write struct members and abstracting from the underlying data storage. 
We propose delegating all conversions and view construction to the C/C++ compiler. 
The compiler receives instructions by the programmer through C/C++ annotations to construct and use views, handling all complexity behind the scenes. 
These annotations apply exclusively to local blocks of (possibly nested) loops.

%
%
A \emph{guided} conversion \cite{Jubertie:2018:DataLayoutAbstractionLayers,Majeti:2014:CompilerDrivenDataLayoutTransformations,Xu:2014:SemiAutomaticComposition} through annotations differs from completely manual (user-driven) or automatic workflows.
The thorny challenge of finding good heuristics when to convert remains out of scope.
We leave the decision ``when'' to convert with the developer but hide the ``how''.
Our \emph{automatic} conversion through compiler techniques ensures that \added[id=copy]{developers can write their algorithms in a memory layout-agnostic way, and the underlying memory layout can even differ from context to context.
C++ template} \deleted[id=Ours]{meta} \added[id=copy]{programming combined with specialised containers would be an alternative to achieve this
\cite{Homann:2018:SoAx,Reinders:2016:XeonPhi,Jubertie:2018:DataLayoutAbstractionLayers,Springer:2018:SoALayout,Strzodka:2011:AbstractionSoA}.
Yet, most of these techniques provide a \emph{static, global} approach. 
The data layout of a struct remains fixed over time.}
Our \replaced[id=Ours]{\emph{temporary, local}}{temporary, local} conversion---where the data layout changes only over a loop's lifespan---frees us from complicated storage state tracking. 
Similar \added[id=copy]{dynamic data structure transformations within a code have been used by several groups successfully \cite{Che:2011:Dymaxion,Gallard:2020:ExaHyPEVectorisation,Majeti:2014:CompilerDrivenDataLayoutTransformations,Vikram:2014:LLVM}.
Our contribution is that we clearly separate storage format considerations from programming, and move all data conversions into the compiler.
}
We accomplish this without requiring developers to alter their original baseline code.

Guided temporary and local views could also be realised through C++ reflections~\cite{Childers:2024:CPPReflections}:
In theory, compiler optimisation passes could use reflection information to reorder data out-of-place, i.e.~copying into a temporary buffer similar to our technique, narrow down these conversions only to data used (dead data access elimination), and ultimately vectorise aggressively.
However, it remains uncertain whether compilers will be able to offer such a complex, multi-step code transformation solely through reflections in the near future.
At the moment, there are, to the best of our knowledge, no automatic view constructions available.
Additionally, reflections require code rewrites, contradicting our goal of minimal invasiveness.
Our approach preserves existing code structure.

Implemented as compiler passes~\cite{Majeti:2014:CompilerDrivenDataLayoutTransformations,Xu:2014:SemiAutomaticComposition}, our approach maintains the code's syntax.
We do not increase the syntactic complexity of the code.
We advance beyond preparatory work \cite{Radtke:2024:AoS2SoA} by identifying views automatically without user interaction. 
Finally, our technique applies not only to genuine AoS arrays but also to scattered data like arrays of pointers to structs.
A guided transformation gathering scattered data into continuous buffers is essential for GPU offloading.

%
%
\added[id=copy]{
We demonstrate the potential of our concepts by means of selected compute kernels from a larger smoothed particle hydrodynamics (SPH) code~\cite{Schaller:2024:Swift}.
Individual SPH interactions can either be strongly memory-bound or rely on compute-intense kernels, while SPH notoriously has to reorder and exchange particles between nodes.
Since we allow the code base to stick to AoS overall, we do not negatively impact algorithmic phases such as the particle boundary exchange or any sorting}, but would expect the automatic conversions to pay off for the compute-intense kernels dominating the total runtime.
Demonstrating that this holds in many cases
\added[id=copy]{
challenges the common knowledge that data conversions pay off only for Stream-like \cite{McCalpin:1995:Stream} kernels or large arrays \cite{Homann:2018:SoAx,Strzodka:2011:AbstractionSoA}.
It also is not in agreement with} \added[id=R1]{with claims that} {temporary data reordering prior to loops or computational kernels is always problematic~\cite{Hundt:2006:StructureLayoutOptimisation,Intel:MemoryLayoutTransformations}.
It can yield significant speedups.
}
As such, our studies and attribute suggestions are relevant for many codes beyond SPH, but they notably provide a protoype how future C/C++ annotations can change the character of performance optimisation, where more code technical refeactorings \cite{Fowler:2019:Refactoring} are deployed into the translation chain freeing human time for analysis and high-level tasks.
This is in the spirit of C++ annotations.

The remainder is organised as follows:
We start with a use case (Section~\ref{section:demonstrator}) that motivates our work but also allows us to introduce our ideas from a developer's perspective. 
Next we introduce our code annotations (Section~\ref{section:annotations}), we discuss their semantics, and we formalise key concepts such that our work can have an impact of future compiler developments (Section~\ref{section:formalism}). 
In Section~\ref{section:realisation-variants}, the annotations are used to implement the core use case ingredients in different ways, 
before we describe the realisation of the annotations in our compiler prototype (Section~\ref{section:realisation}).
We continue with case studies on their impact in Section~\ref{section:results}.
A brief summary and outlook in Section~\ref{section:conclusion} close the discussion.

\section{Case study}
\label{section:demonstrator}

We study our ideas through a smoothed particle hydrodynamics (SPH) code written in C++.
SPH describes fluid flow using moving particles that carry distributions such as velocity and physical properties, thereby representing the underlying fluid dynamics.
The method finds applications in a variety of application
domains~\cite{Lind:2020,Price:2012:SPH}, notably in computational astrophysics (see for example {\sc
gadget}-2~\cite{Springel-g2:2005} or SWIFT~\cite{Schaller:2024:Swift}).

\subsection{Algorithm blueprint}

SPH's particles move in space through an explicit time-stepping---we employ leapfrog---and exchange quantities between each other. 
SPH derives these particle-particle interactions through integral formulations over the underlying partial differential equations, where quantities of interest are approximated through functions with local support. 
We employ the quartic spline (M5) kernel~\cite{Monaghan:1985:Kernel}, which can be seen as prototype for more complex interaction kernels~\cite{Denhen:2012:SPHConvergence}.

The core compute steps of per time step of a baseline algorithm for any SPH code read as follows:

\begin{enumerate}
  \item \added[id=copy]{We determine the smoothing length of each particle, i.e.~the
  interaction area (circle or sphere around a particle) as well as its \emph{density}. Per particle, this initial step studies all
  nearby neighbouring particles, computes the local density, and finally decides} \replaced[id=R1]{whether}{if} \added[id=copy]{to shrink
  or increase the interaction radius.} Changing the radius triggers a 
  recomputation of the density, as we now have to take more or fewer particles into
  account, i.e.~the process repeates iteratively until each particle reaches
  its termination criterion.
  \item \added[id=copy]{We compute the \emph{force} acting on each particle. The force
  is the sum over all forces between a particle and its neighbours within the interaction radius.}
  \item We \emph{kick} the particle by half a time step, i.e.~we accelerate it.
  \item We \emph{drift} a particle, i.e.~update its position.
  \item We \emph{kick} a second time, i.e.~add
  another acceleration. This step also resets various temporary properties of each particle, preparing them for the next density calculation.
\end{enumerate}


\noindent
\added[id=copy]{
We focus exclusively on SPH with hydrodynamics, i.e.~ignore long-term
interactions such as gravity.
More complex physical models start from this baseline algorithmics \cite{Schaller:2024:Swift}.
Consequently, our particles have a small, finite
interaction radius---which may change over time, but is always small---and the arising interaction matrix is sparse and localised:
Let $N$ be the particle count.
The force and density calculation are in $\mathcal{O}(N^2)$ yet iterate solely 
over their neighbourhood.
The kicks and drifts update a particle without any particle--particle interaction and hence are genuinly in $\mathcal{O}(N)$.
}

Like most major SPH codes, our work employs a mesh helper structure where particles are sorted a priori into cells.
The cells are chosen such that two particles may interact if and only if they reside within neighbouring cells.
This reduces the cost of the force and density calculations:
Rather than looping over all particles in the domain, we loop over cells, launching one \emph{compute kernel} per cell.
This kernel in turn loops over the particles within the cell and examines only those neighbouring particles that are bucketed into vertex-connected neighbour cells.
We \replaced[id=R1]{continue}{contine} to have an $\mathcal{O}(N^2)$ algorithm locally, but $N$ is the upper bound on the number of particles per cell.
It is way smaller than the total particle count.

In our code base, we employ a spacetree formalism \cite{Weinzierl:2015:pidt} to create the mesh and sort the particles into the leaves of the tree.
The tree facilitates dynamic mesh refinement and coarsening, which in practice keeps $N$ per cell close to constant for most cells\replaced[id=R1]{,}{.}
$N \approx ppc$ (particles per cell).
Further to that, using a tree spawning cells allows us to implement domain decomposition straightforwardly:
Mesh cells are distributed over nodes and threads, and all particles follow the ownership of the cells they are sorted into.
Consequently, we have to maintain a set of ghost cells and ghost particles:
Each local cell must access all neighboring cells and their particles for force and density calculations.
If a particle sits in a neighbouring cell not assigned to the local node, we have to maintain a copy of this particle locally, and update its properties after each algorithm step.

Any interacting particle pair must occupy neighbouring cells.
Sorting particles into the mesh and the mesh management itself are not free.
Therefore, codes tend to enlarge cells bigger, increasing $N$.
This raises computational load per kernel call, but overall reduces compute time.
It remains a dark art, i.e.~empiric knowledge what reasonable cell sizes and particle counts are.

\subsection{Particle memory layout and data flow}

We model individual particles as structs.
This naturally translates the physical model into software. 
The resulting AoS storage benefits particle sorting into the mesh: when particles move, we must update cell membership. 
Moving and sorting are comparatively inexpensive operations in our code base, but they synchronise all other compute steps:
As long as not all particles are in the right place, we cannot trigger any follow-up computations.
Therefore, we use a struct-centric data model that ensures minimal data movements, as we can copy whole structs at once, and makes particle movements as well as halo data exchange simple.
Indeed, our codebase stores particle structs on the heap and uses lists of pointers within the mesh to index these scattered memory regions, i.e.~we employ an array of structs (AoS) with pointer indirection.

The structs modelling particles in our code can hold from a few quantities up to hundreds of doubles, depending on the physical model.
Our benchmark code works with a default memory footprint of 272 bytes per particle.
Some of these bytes encode administration information, others store physical
properties.
For many steps, only few data members enter the equations implemented by the compute kernels.
The density calculation starts from the density and smoothing length of the
previous time step and updates those two quantities plus some others such as the neighbour count, the rotational velocity vector, and various derivative properties.
Smoothing length, density and further properties then feed into the force
accumulation which eventually yields an acceleration per particle.
Kicks are relatively simple, i.e.~add scaled accelerations onto the velocity.
The second kick in our implementation also resets a set of predicted values
which feed into the subsequent density calculation.
Drifts finally implement a simple Euler time integration step, i.e.~a tiny
daxpy.

Although pointer-indexed AoS is our primary modeling technique, we can instruct the particle sorting to ensure continuous memory storage of particles within cells. 
It then moves around data throughout any resort ensuring that all particles of one cell end up in one continuous chunk of memory.
Consequently, we can loop through the particles within one cell with a normal loop enabling coalesced memory access.
For the $\mathcal{O}(N^2)$ algorithms, we can loop over chunk pairs of particles.

\subsection{Vanilla compute kernel version}

\begin{lstlisting}[language=C++,
                   label=algorithm:demonstrator:blueprint,
                   caption=Schematic illustration of the density loop.,
                   basicstyle=\footnotesize]
void ForAllParticlePairs(list<Particle*> &localParticles,
                         list<Particle*> &activeParticles) {
  for (auto *localParticle : localParticles) {
    for (auto *activeParticle : activeParticles) {
      if (DensityPredicate(localParticle, activeParticle)) {
        Density(localParticle, activeParticle);
      }
    }
  }
}
\end{lstlisting}

We discuss our baseline implementation, i.e.~the starting point of any optimisation, by means of the density loop (Listing~\ref{algorithm:demonstrator:blueprint}), which is schematically the same as the force loop. 
Kicks and drifts lack the outer loop and therefore are simpler.

\begin{figure}
 \begin{center}
  \includegraphics[width=0.25\textwidth]{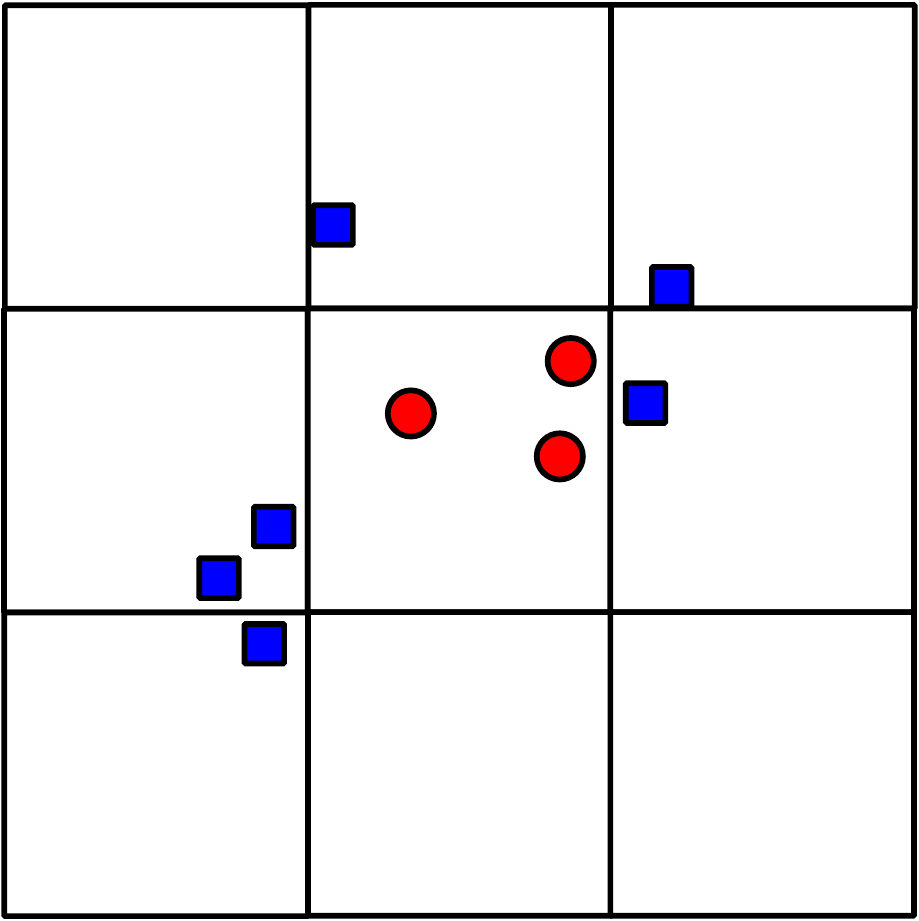}
 \end{center}
 \caption{
  A compute kernel updates all local particles (red circles) within a cell.
  These particles depend upon their neighbours within the cell plus particles hosted by neighbouring cells (blue squares).
  The set of potentially influencing particles---if particles really interact is subject to a particle-specific interaction radius---is named active set and yields a superset of the local particles.
  \label{figure:demonstrators:active-vs-local}
 }
\end{figure}

In our code, we distinguish active from local particles, with local particles being all particles within a cell and active particles all potential interaction partners (Figure~\ref{figure:demonstrators:active-vs-local}). 
The local particles therefore are a subset of the active particles.
As we have an $\mathcal{O}(N^2)$ algorithm per cell, we see that the code becomes a nested \texttt{for}-loop.
Each loop runs over a C++ container which hosts pointers to the particles; as long as we make no further assumptions on the particle storage order.

Semantically, it makes no difference whether we loop over the active or local particles in the outer loop.
We may permute them.

The actual semantics, i.e.~physics, are injected through two function calls (\texttt{DensityPredicate} and \texttt{Density}).
The latter contains all updates to particles, while the predicate ensures that we update the right particles:
For example, it checks that only particles within the interaction radius are taken into account,
and that we avoid self-interaction.


\subsection{Manually tweaked compute kernel}

\algrenewcommand\algorithmicindent{0.4em}%

\begin{lstlisting}[language=C++,
                   label=algorithm:demonstrator:blueprint-rewritten,
                   caption=Manual rewrite of the blueprint code from Listing~\ref{algorithm:demonstrator:blueprint} into a temporal local variant.,
                   basicstyle=\footnotesize]
void ForAllParticlePairs(list<Particle*> &localParticles,
                         list<Particle*> &activeParticles) {
  double *soa_lParticle_density = new double[localParticle.size()];
  double *soa_lParticle_pressure = new double[localParticle.size()];
  // ... declare buffers for local and active sets
  
  int idx = 0;
  for (auto *particle : localParticles) {
    soa_lParticle_density[idx] = particle->getDensity();
    soa_lParticle_pressure[idx] = particle->getPressure();
    // ... copy data from AoS to SoA buffers
    idx++;
  }
  
  for (int iLoc=0; iLoc<localParticles.size(); iLoc++) {
    for (int iAct=0; iAct<activeParticles.size(); iAct++) {
      if (DensityPredicate(/* operate on SoA data */)) {
        Density(/* operate on SoA data */);
      }
    }
  }
  
  idx = 0;
  for (auto *particle : localParticles) {
    particle->setDensity(soa_lParticle_density[idx]);
    particle->setPressure(soa_lParticle_pressure[idx]);
    // ... copy data from SoA to AoS buffers
    idx++;
  }
  
  delete[] soa_lParticle_density;
  delete[] soa_lParticle_pressure;
  // ... clean up buffers
}
\end{lstlisting}

\added[id=copy]{Converting AoS into SoA is an evergreen in high-performance computing,
once we have committed to AoS as development data structure (cmp.~for example}
\cite{Gallard:2020:Roles,Gallard:2020:ExaHyPEVectorisation,Homann:2018:SoAx,Hundt:2006:StructureLayoutOptimisation,Majeti:2014:CompilerDrivenDataLayoutTransformations,Intel:MemoryLayoutTransformations,Jubertie:2018:DataLayoutAbstractionLayers,Reinders:2016:XeonPhi,Springer:2018:SoALayout,Strzodka:2011:AbstractionSoA,Sung:2012:DataLayoutTransformations,Vikram:2014:LLVM,Xu:2014:SemiAutomaticComposition}).
In the present example, a programmer would insert a memory allocation to hold arrays of the input data (Listing~\ref{algorithm:demonstrator:blueprint-rewritten}).
The term SoA suggests that these arrays are packaged into a struct of its own.
We omit this packaging here, i.e.~we work with arrays directly.
The signatures of the two core routines have to be amended to operate on the temporary data.
Finally, we have to convert the temporary SoA data \replaced[id=R1]{back into}{backinto} AoS, i.e.~update the original.

\begin{concept}
  In a \emph{temporary} reformulation, the data is only kept for a well-defined, short time in SoA format.
\end{concept}

\begin{concept}
  A \emph{local} rewrite converts into the SoA format over a well-defined building block such as a loop. If it ripples through multiple functions, it only propagates along the call tree.
\end{concept}

\noindent
While not challenging in itself, rewriting the code is obviously tedious and error-prone.
Therefore, we propose to let a compiler realise this rewrite:

\begin{concept}
  A \emph{guided} AoS-to-SoA conversion requires the programmer to explicitly instruct the compiler for which code blocks to convert the data into SoA.
\end{concept}

\noindent
Any programmer rewriting AoS manually into SoA naturally only converts those members of a struct into an array or vice versa that are actually used later on or have changed, respectively:
As the density calculation reads the positions of a particle yet never alters them, the final update loop would not mirror back the positions into the original data structure.
Our work proposes to deploy all of these steps and considerations to a tool, i.e.~the compiler:

\begin{concept}
  An \emph{automatic} conversion deploys the task to identify which members of a struct need converting into an array to a tool, i.e.~no user interaction is required. Furthermore, it also generates all the conversion routines. 
\end{concept}

\noindent
Previous work referred to this minimal subset of a struct over a code block as a view \cite{Radtke:2024:AoS2SoA}.
Other literature refers to this as peeling \cite{Hundt:2006:StructureLayoutOptimisation}.
We generalise the notion of a view in our work:

\begin{concept}
 A \emph{view} is a logical concept over a struct. A view knows which members of this struct are read, which are written and how they are stored in a given region of a program.
\end{concept}

\noindent
Working with a view means that code can be written over structs or pairs of structs, respectively, even though the underlying data structure is SoA.
The view ought to hide the memory layout details from the rest of the code.
It also hides \deleted[id=R1]{the} that only some struct members are read or written and hence converted into another ordering.
With views, the core routines \texttt{DensityPredicate} and \replaced[id=R2]{\texttt{Density}}{\textsc{Density}} do not have to be altereed even if the compiler decides to change the underlying memory layout.

Our manual conversion illustrates that temporary, local conversion into SoA incurs unavoidable data movement costs. 
The implicit hope that motivates these conversions is that speedups---achieved through improved vectorisation within loop bodies or more efficient cache use in nested loops---will compensate for this overhead and yield overall faster execution.

\section{Source code annotations}
\label{section:annotations}

\begin{lstlisting}[language=C++,
                   label=algorithm:annotations:example,
                   caption=Annotated blueprint code. GPU offloading taking data conversions and views into account is enabled by adding \texttt{[[clang::soa\_conversion\_offload\_compute]]} to the source code right before the very first \texttt{for} loop.,
                   basicstyle=\footnotesize]
void ForAllParticlePairs(list<Particle*> &localParticles,
                         list<Particle*> &activeParticles) {
  [[clang::soa_conversion]]
  for (auto *localParticle : localParticles) {
    [[clang::soa_conversion_hoist(1)]]
    for (auto *activeParticle : activeParticles) {
      if (DensityPredicate(localParticle, activeParticle)) {
        Density(localParticle, activeParticle);
      }
    }
  }
}
\end{lstlisting}

%
%
\added[id=copy]{We propose that developers stick to AoS as modelling data structure, and request temporary data transformations into SoA that are executed automatically by the compiler.
For this we introduce new C++ attributes}
(Listing~\ref{algorithm:annotations:example}):

\begin{itemize}
  \item The attribute \texttt{[[clang::soa\_conversion]]} must precede a loop over a C++ container and instructs the translator to construct a view over this container that converts all read (in) and all written data members (out) into a SoA.
  \item The attribute \texttt{[[clang::soa\_conversion\_hoist(n)]]} can be applied to an inner loop. It has the same semantics as a simple conversion yet moves the conversion $n$ levels out of the loop hierarchy.
  \item The attribute \texttt{[[clang::soa\_conversion\_compute\_offload]]} deploys the loop execution, subject to the conversions, onto OpenMP's target offloading.
\end{itemize}

\subsection{Annotation semantics}
Each annotation triggers a temporary, local, automatic rewrite of the container indexed by the loop into SoA.
\added[id=copy]{We instruct the compiler to perform a temporary out-of-place transformation behind the scenes} \cite{Sung:2012:DataLayoutTransformations}:

The annotation makes the compiler analyse all data accesses to the container data within the loop body. 
Let the struct members read be described by a set $\mathbb{A}_{\text{in}}$,
while data members written end up in $\mathbb{A}_{\text{out}}$.
\added[id=copy]{
The \texttt{soa\_conversion} statement then adds a prologue to the loop.
}
This prologue creates one logical (temporary) array per entry in $\mathbb{A}_{\text{in}} \cup \mathbb{A}_{\text{out}}$.
After that, the entries from the container are copied over into the temporary data structures for all entries in $\mathbb{A}_{\text{in}}$.
$\mathbb{A}_{\text{out}} \setminus \mathbb{A}_{\text{in}} $ does not require any preparation.
Next, the compiler ensures that the transformed loop accesses the SoA buffers instead of the \deleted[id=Ours]{original} AoS data. 
Finally, we synchronise the temporary data with the origin data structure.
This is achieved through an epilogue following the main computation loop.
The epilogue step loops over the container and moves the temporary data indexed by $\mathbb{A}_{\text{out}}$ back into the original data structure.
Once this epilogue terminates, any changes induced by the loop body on the view are re-synchronised with the original loop container.
Eventually, we free the temporary data, i.e.~we destroy the view.

\subsection{Annotation properties and constraints}

To make the automatic construction of correct views work, the compiler analysis has to track all data accesses of the AoS data inside the loop body, including accesses inside functions called from within the loop body. 
All definitions of the functions called, either directly or indirectly, need to be known. 

We assume that the loop body is atomic, i.e.~free of side effects caused outside of the loop body.
Aliasing effects would fall into this rubric but also accesses to atomic variables through multiple threads:
As we construct the view through an out-of-place transformation, any access to an entry of the loop container from outside the loop, even though correctly declared as atomic, will continue to affect the original container element, while the loop works on its working copy.
The change is not visible to the loop body and will eventually be overwritten.
Concurrent access to a container property within the loop body is not a problem.
In this case, all concurrent \replaced[id=R1]{accesses are}{access is} encapsulated.

The concurrency statements hold exclusively for data members from $\mathbb{A}_{\text{in}} \cup \mathbb{A}_{\text{out}}$.
The view splits the members of the struct into two categories:
While the loop is running, the data items from $\mathbb{A}_{\text{in}} \cup \mathbb{A}_{\text{out}}$ within the container become temporarily invalid.
The remaining struct members however remain valid as the epilogue's mapping from the transformed SoA data onto the original container is, in general, not surjective.

\subsection{Data replication due to aliasing}

For struct members in $\mathbb{A}_{\text{in}} \cap \mathbb{A}_{\text{out}} \not= \emptyset$, an out-of-place transformation yields only one instance of these data.
We do not create a separate instance for $\mathbb{A}_{\text{in}}$ and one for $\mathbb{A}_{\text{out}}$.
Yet, only those data members in $\mathbb{A}_{\text{out}}$ are eventually picked from the transformed data and copied back.

For nested loops with aliasing, i.e.~nested loops where two containers reference joint data, we have to be more careful.
Let $\mathbb{A}_{\text{in}}^{(outer)},\ \mathbb{A}_{\text{out}}^{(outer)}$ and $\mathbb{A}_{\text{in}}^{(inner)},\ \mathbb{A}_{\text{out}}^{(inner)}$ be the active sets of the outer or inner loop, respectively.
If the underlying containers index distinct sets of objects, no further discussion is required.
However, ambiguity problems can arise if 

\[
\mathbb{A}_{\text{out}}^{(inner)} \cap \left( \mathbb{A}_{\text{in}}^{(outer)} \cup \mathbb{A}_{\text{out}}^{(outer)} \right) \not= \emptyset.
\]

\noindent
In this case, the outer loop's prologue will convert \replaced[id=R1]{$\mathbb{A}_{\text{in}}^{(outer)} \cup \mathbb{A}_{\text{out}}^{(outer)}$}{$\mathbb{A}_{\text{out}}^{(outer)} \cup \mathbb{A}_{\text{out}}^{(outer)}$} into a temporary new data set.
The inner loop's prologue will also create a new dataset, effectively duplicating the overlapping data.
As we have a bitwise replica of all data items, the \replaced[id=R1]{replicas}{replica} are consistent.
The subsequent epilogue of the inner loop now modifies the members identified through $\mathbb{A}_{\text{out}}^{(inner)}$.
However, these modifications are not mirrored in the set \replaced[id=R1]{$\mathbb{A}_{\text{in}}^{(outer)} \cup \mathbb{A}_{\text{out}}^{(outer)}$}{$\mathbb{A}_{\text{out}}^{(outer)} \cup \mathbb{A}_{\text{out}}^{(outer)}$}, i.e. the two datasets become inconsistent.

We label setups with ambiguous conversion outcomes as undefined or corrupted, i.e.~we do not support them.
Indeed, our SPH code's distinction of active vs.~local sets does yield overlapping data, i.e.~aliasing effects, but it remains well-defined, as we always write exclusively to the local data and furthermore ensure in our implementation that these updated data are never read from the active data.

\subsection{Offloading through OpenMP}

OpenMP's \texttt{target} clause and its offloading infrastructure makes it straightforward to deploy compute kernels to an accelerator.
Offloading however requires a proper mapping of the CPU's memory, i.e.~we have to ensure a correct memory allocation on the device and efficient data transfers between the host and the accelerator.
The existing SoA transformation classifies memory access over the struct members into three types: 
read-only ($\mathbb{A}_{\text{in}} \setminus \mathbb{A}_{\text{out}}$), write-only ($\mathbb{A}_{\text{out}} \setminus \mathbb{A}_{\text{in}}$), and read-write ($\mathbb{A}_{\text{in}} \cap \mathbb{A}_{\text{out}}$). 
They directly define which memory mappings or allocations are required if we work with a distributed memory model:
Read-only data is allocated on the device, and data is transferred from the host to the device before the kernel execution. 
Write-only buffers are allocated on the device, and data is transferred from the device to the host after kernel execution. 
Finally, read-write data is allocated on the device, with two memory transfers taking place---one from the host to the device before execution and another from the device to the host after kernel execution.
In a shared memory model, all memory movements are delegated to the underlying hardware/driver/runtime stack.

Our annotation exclusively relates to the iteration's container. If additional data, such as static variables, need to be mapped to the device, we delegate the responsibility of mapping it correctly to the user.

For the realisation of the loop iteration, we assume that we can directly exploit \texttt{teams distribute parallel for}, i.e.~that the underlying loop iterations are embarassingly parallel.
For an annotation prototype, this is sufficient.
\added[id=Ours]{
 However, such a high level of abstraction strips us from the opportunity to apply any fine tuning:
 GPU kernels typically deliver high performance if and only if we carefully balance between team distribution grain sizes and thread counts.
}
In the long term, a closer integration into OpenMP's syntax---compare the use of attributes in the OpenMP Technical Report 8 \cite{OpenMP:TechnicalReport8}---\added[id=Ours]{therefore} would be beneficial.

\section{Formal code transformations and algorithm description}
\label{section:formalism}

Let $D = [S_0,S_1,S_2,\ldots]$ be an array of objects, i.e.~instances of structs.
As we store data as AoS, the memory holds $[S_0.a_0,S_0.a_1,S_0.a_2,\ldots,S_1.a_0,S_1.a_1],\ldots$, where the data members $a_i$ span the set $\mathbb{A}$.
$ D \gets f(D)$ represents the loop body of interest.
Function and surrounding loop together define the overall compute kernel.
Without loss of generality, we encounter

\begin{eqnarray*}
  S_j = [S_j.a_0, S_j.a_1, \ldots] & \gets & f([S_j.a_0, S_j.a_1, \ldots]) \qquad \text{and} \\
  S_j = [S_j.a_0, S_j.a_1, \ldots] & \gets & \sum _{i\not=j} f([S_j.a_0, S_j.a_1, \ldots],[S_i.a_0, S_i.a_1, \ldots])
\end{eqnarray*}

\noindent
for our compute kernels of linear complexity or in $\mathcal{O}(N^2)$, respectively.
The loop body itself hosts exclusively unary and binary operations.

\subsection{Data subset construction}

Let $f$ of a loop body be completely visible, i.e. no external functions are called, either directly or indirectly, in the loop body.
We construct the sets $\mathbb{A}_{\text{in}} \subseteq \mathbb{A}$ and $\mathbb{A}_{\text{out}} \subseteq \mathbb{A}$ by running through the loop body line by line and inspecting all read/write operations on the AoS data.
For any operation

\begin{equation}
 x \gets \psi(y_0,y_1,\ldots)
 \label{equation:formalism:assignment}
\end{equation}

\noindent
with $y_i \in \mathbb{A}$ from the traversal container, we add the data members to the set 
\[
 \mathbb{A}_{\text{in}} \gets \mathbb{A}_{\text{in}} \cup \{ y_i \}.
\]

\noindent
Analogously, $x$ is added to the out set if it refers to one of the data members of the container:
\[
 \mathbb{A}_{\text{out}} \gets \mathbb{A}_{\text{out}} \cup \{ x \} \quad \text{if} \quad x \in \mathbb{A}.
\]

\noindent
In practice, a less strict representation of the loop body can be used, i.e.~we do not have to restrict outselves to expressions of the form (\ref{equation:formalism:assignment}).

Let $f$ invoke a subroutine $f_{\text{sub}}$ with $f_{\text{sub}}$ being available (visible), too.
In this case, we can recursively apply our analysis to $f_{\text{sub}}$ to construct a $\mathbb{A}_{\text{in}}^{f_{\text{sub}}}$ and a $\mathbb{A}_{\text{out}}^{f_{\text{sub}}}$.
Once complete, we fuse the sets

\begin{eqnarray}
\mathbb{A}_{\text{in}} & \gets & \mathbb{A}_{\text{in}} \cup \mathbb{A}_{\text{in}}^{f_{\text{sub}}}
\quad \text{and} 
\label{equation:formalism:union-over-sets}
\\
\mathbb{A}_{\text{out}} & \gets & \mathbb{A}_{\text{out}} \cup \mathbb{A}_{\text{out}}^{f_{\text{sub}}}.
\nonumber
\end{eqnarray}

\noindent
We exhaustively traverse the call graph.
Recursive or re-occuring function calls are covered by this formalism once we recursively analyse if and only if the resulting union sets in 
(\ref{equation:formalism:union-over-sets}) continue to grow.

\subsection{Operations}

Let $N_{\mathbb{A}_{\text{in}} \cup \mathbb{A}_{\text{out}}}$ be the operation that narrows down the struct members to the ones read and written by a loop body.
For example, if $f$ reads $\mathbb{A}_{\text{in}} = \{ S_i.a_7 \}$ only and writes into $\mathbb{A}_{\text{out}} = \{S_i.a_3\}$, then 
\[
  [S_i.a_3,S_i.a_7] = N_{\mathbb{A}_{\text{in}} \cup \mathbb{A}_{\text{out}}}([S_i.a_0,S_i.a_1,\ldots])
\]

\noindent
for a single struct.
We can canonically apply it to a sequence of objects.

The operation $W_{\mathbb{A}_{\text{out}}}$ widens the object again. 
We use the notation such that 

\[
  D = \left( W_{\mathbb{A}_{\text{out}}} \circ N_{\mathbb{A}_{\text{in}} \cup \mathbb{A}_{\text{out}}} \right) (D),
\]

\noindent
and our optimisation is based upon the idea that 

\begin{equation}
  f(D) = \left( W_{\mathbb{A}_{\text{out}}} \circ f \circ N_{\mathbb{A}_{\text{in}} \cup \mathbb{A}_{\text{out}}} \right) (D),
    \label{equations:narrowing-widening}
\end{equation}

\noindent
i.e.~that we can pick out all the read and written data from $D$, apply (a rewritten) $f$ to these temporary values, and eventually mirror the changes back into the container holding $D$.
Let

\begin{eqnarray}
  \lbrack S_0.a_0, S_1.a_0, S_3.a_0, \ldots , S_1.a_0, \ldots \rbrack 
    & \gets & 
    C_{AoS}^{SoA}( D ) 
    \label{equations:AoStoSoA}
    \\
  \lbrack S_0.a_0,S_0.a_1,S_0.a_2,\ldots,S_1.a_0,\ldots \rbrack 
    & \gets & (C_{SoA}^{AoS} \circ C_{AoS}^{SoA})( D ). \nonumber 
\end{eqnarray}

\noindent
formalise the conversion from AoS into SoA with 
$ C_{SoA}^{AoS} = \left( C_{AoS}^{SoA} \right)^T$.
The realisation of $f$ over the whole array can now be broken down into three steps:
A conversion into SoA, then the explicit evaluation of the loop body, followed by a conversion back into AoS.
The notion of a view combines these operations (\ref{equations:narrowing-widening}) and (\ref{equations:AoStoSoA}) into

\begin{eqnarray}
  f(D) = \left( W_{\mathbb{A}_{\text{out}}} \circ C_{SoA}^{AoS} \circ \hat f \circ C_{AoS}^{SoA} \circ N_{\mathbb{A}_{\text{in}} \cup \mathbb{A}_{\text{out}}} \right) (D).
  \label{equation:view}
\end{eqnarray}

\noindent
The right-hand side replaces the original user code $f(D)$.
While permuting the narrowing and SoA conversion in (\ref{equation:view}) would preserve semantics, it makes no sense from an implementation perspective.
In practice, these two functions collapse into a single operation requiring memory gathering, and we would always only work on the data ``left over'' by the narrowing:
it is unreasonable first to convert all data into SoA and then to pick from the outcome.
The epilogue $W_{\mathbb{A}_{\text{out}}} \circ C_{SoA}^{AoS}$ introduces scattered memory accesses.
Again, it makes no sense to permute the operators.
Indeed, widening is not properly defined over data formats of differing types.
While $C_{SoA}^{AoS}$ and $\left( C_{SoA}^{AoS} \right) ^T$ can be realised in-situ, the conversion in combination with the narrowing and widening requires us to hold the converted data in some explicit storage $\hat D$.

\subsection{GPU offloading}

If we deploy a loop onto the accelerator with memory address translation as we find it in OpenMP, we either have to copy the data onto the accelerator prior to the loop or we rely on shared memory.
We obtain

\begin{eqnarray}
  f(D) = \left( W_{\mathbb{A}_{\text{out}}} \circ C_{SoA}^{AoS} \circ M_{GPU}^{CPU} \circ \hat f_{GPU} \circ M_{CPU}^{GPU} \circ C_{AoS}^{SoA} \circ N_{\mathbb{A}_{\text{in}} \cup \mathbb{A}_{\text{out}}} \right) (D),
  \label{equation:gpu}
\end{eqnarray}

\noindent
where $M_{GPU}^{CPU}$ moves data from the host to the GPU.
We explicitly use OpenMP routines.
If we use shared memory, these moves are implicitly triggered by the hardware.
We can omit $M_{GPU}^{CPU}$ and $M_{CPU}^{GPU}$.

$M_{CPU}^{GPU} = \left( M_{GPU}^{CPU} \right)$ holds if and only if $\mathbb{A}_{\text{in}} = \mathbb{A}_{\text{out}}$.
Otherwise, each operation works only on its relevant subset.
We note that $M_{CPU}^{GPU} \circ C_{AoS}^{SoA} \circ N_{\mathbb{A}_{\text{in}}}$ provides space for optimisation, as we can parameterise the device moves over the narrowed set:
Only data in $\mathbb{A}_{\text{in}}$ has to be copied to the accelerator, while we can allocate $\mathbb{A}_{\text{out}} \setminus \mathbb{A}_{\text{in}}$ on the accelerator without any additional data movements.

Given the observation that memory movements become a limiting factor in GPU utilisation, the order of operations in (\ref{equation:gpu}) is natural.
However, it is not clear that this order is optimal in all cases.
For very tightly integrated future GPU-CPU solutions,

\begin{eqnarray}
  f(D) & = & \left( M_{GPU}^{CPU} \circ W_{\mathbb{A}_{\text{out}}} \circ C_{SoA}^{AoS} \circ \hat f  \circ C_{AoS}^{SoA} \circ N_{\mathbb{A}_{\text{in}} \cup \mathbb{A}_{\text{out}}} \circ M_{CPU}^{GPU} \right) (D)
  \qquad \text{or}
  \nonumber \\
       & = & \left( W_{\mathbb{A}_{\text{out}}} \circ M_{GPU}^{CPU} \circ C_{SoA}^{AoS} \circ \hat f  \circ C_{AoS}^{SoA} \circ M_{CPU}^{GPU} \circ N_{\mathbb{A}_{\text{in}} \cup \mathbb{A}_{\text{out}}} \right) (D)
  \label{equation:formalism:permute-cpu-gpu-conversion}
\end{eqnarray}

\noindent
might be reasonable alternatives to explore---in particular if the AoS--to--SoA conversions can be done in-situ by the GPU \cite{Che:2011:Dymaxion}.

Kernel memory mapping in OpenMP requires three key ingredients: the buffer, its size, and the mapping type, i.e.~in, out or inout. 
$M_{CPU}^{GPU}$ and $M_{GPU}^{CPU}$ have direct access to these three properties.
The buffer's location is given by the container, 
its size is known due to C++ container \texttt{size()}, 
and the mapping type results from the access pattern analysis. 

\section{Kernel realisation variants}
\label{section:realisation-variants}

We have \replaced[id=R1]{several realisation variants}{various variations} of the blueprint from Listing~\ref{algorithm:demonstrator:blueprint} at hand that we can augment with the annotations.

\subsection{Generic kernels}

\begin{figure}[htb]
\centering
 \includegraphics[width=0.75\textwidth]{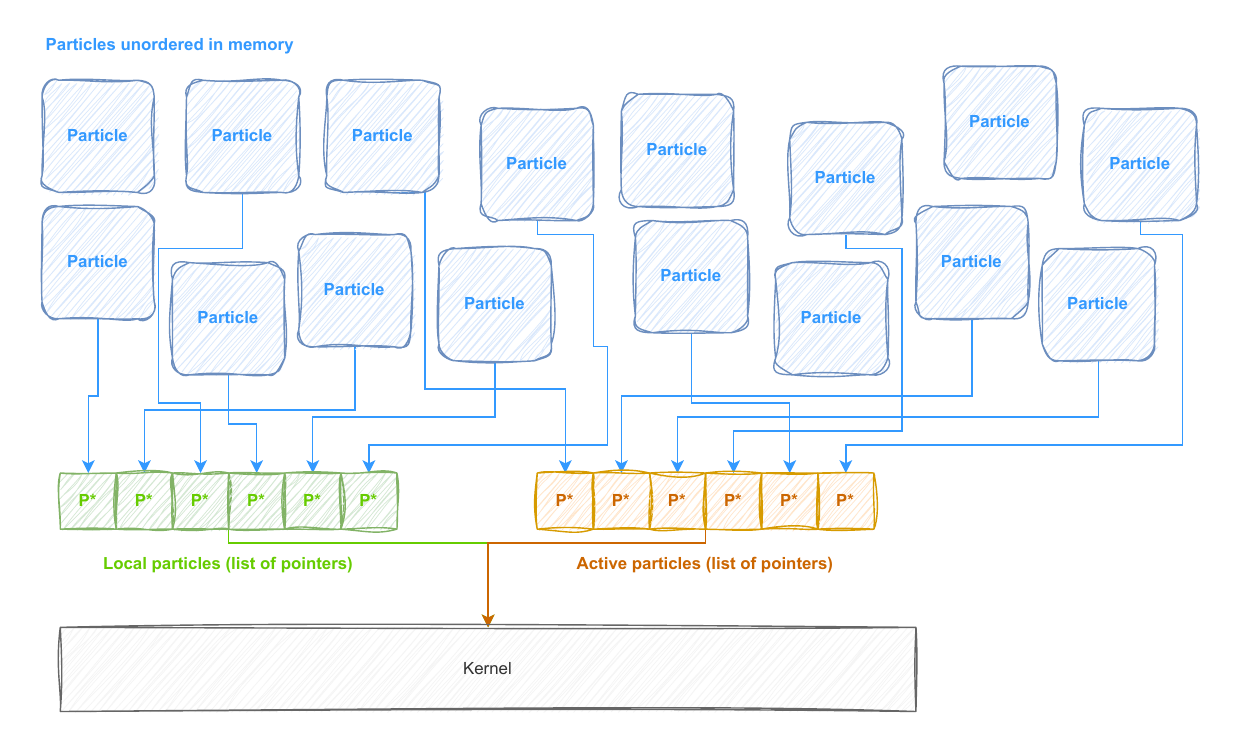}
 \caption{
   The generic kernel is written over arrays of pointers to structs.
   They can be scattered.
   \label{sketches:memlayout:unordered}
 }
\end{figure}

Our generic vanilla version remains the variant that does not exploit any knowledge about the underlying memory arrangement.
It simply loops over lists of pointers, which might point to arbitrarily scattered memory locations (Figure~\ref{sketches:memlayout:unordered}).

\begin{figure}[htb]
\centering
 \includegraphics[width=0.75\textwidth]{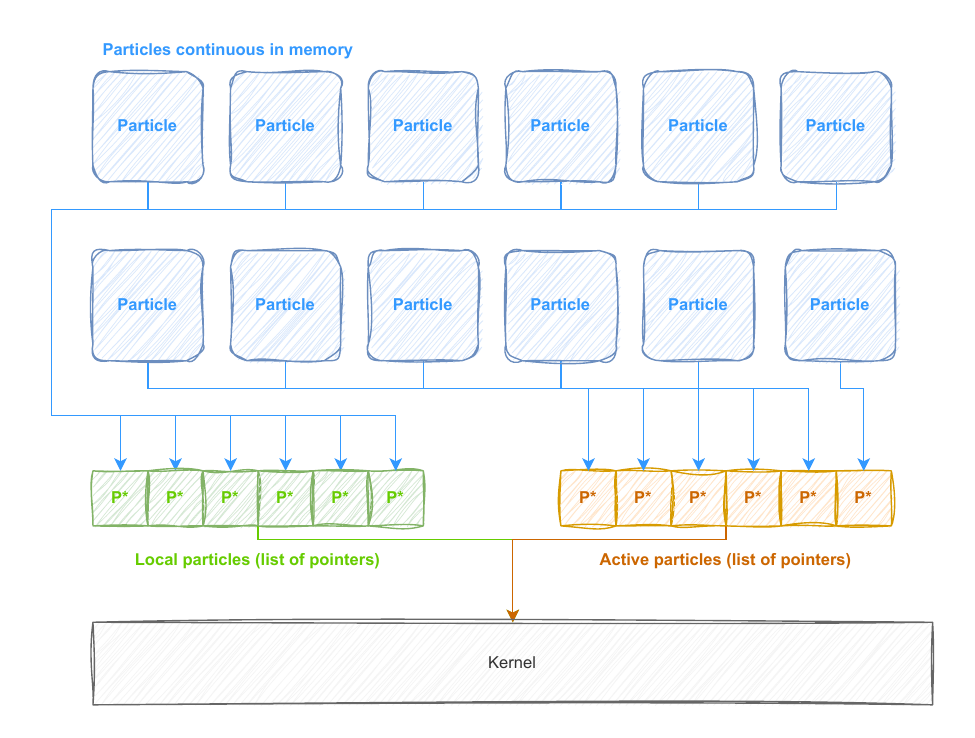}
 \caption{
  Our SPH code base can be instructed to sort the particles after each drift step, such that particles within one cell are stored as one continuous block in memory.
  \label{sketches:memlayout:ordrered}
 }  
\end{figure}

However, we offer the opportunity within the underlying SPH code base to make each particle movement be followed by a particle sorting step that rearranges the memory such that particles from one cell end up in one continuous memory block (Figure~\ref{sketches:memlayout:ordrered}).
This variant in principle facilitates coalescent memory access.
The generic kernel does not exploit such knowledge explicitly, as it works with lists of pointers without any visibility into any underlying allocation patterns, but it should, in principle, benefit from the memory arrangement.

\subsection{Coalesced memory access and loop permutations}

\begin{lstlisting}[language=C++,
                   label=algorithm:kernel-variants:coalesced,
                   caption=Rewrite of the algorithm exploiting the fact that we know that some particle chunks are stored consecutively in memory.,
                   basicstyle=\footnotesize]
void ForAllParticlePairs(list<Particle*> localParticles,
                         list<Particle*> activeParticles,
                         list<int> localParticleChunkSizes,
                         list<int> activeParticleChunkSizes) {
  int globalLPidx = 0;
  for (int lPChunkSize : localParticleChunkSizes) {
    auto *lPChunk = localParticles[globalLPidx];
    for (int lPidx=0; lPidx < lPChunkSize; lPidx++) {
      int globalAPidx = 0;
      for (int aPChunkSize: activeParticleChunkSizes) {
        auto *aPChunk = activeParticles[globalAPidx];
        for (int aPidx=0; aPidx <a PChunkSize; aPidx++) {
         if (DensityPredicate(lPChunk[lPidx], aPChunk[aPidx]))
            Density(lPChunk[lPidx], aPchunk[aPidx]);
        }
      }
      globalAPidx += aPChunkSize;
    }
    globalLPidx += lPChunkSize;
  }
}
\end{lstlisting}

As part of (historic) performance engineering, we provide a second kernel variant that is passed an additional meta data structure indexing the particle sequence (Listing~\ref{algorithm:kernel-variants:coalesced}):
The meta data stores a sequence of integers.
Each integer identifies how many pointers within the AoS input point to consecutively stored objects on the heap.
This \replaced[id=R2]{mirrors}{mirros} index data structures as found in compressed sparse row storage formats \cite{Liu:1986:Compact} for example.

The compute kernel exploiting coalesced memory then does not host one loop over a continuous piece of data for the local and one for the active particles.
Instead, it has two outer loops over chunks of particles.
Within each chunk, it can then exploit a plain for loop over chunks-size consecutive particles.
In our SPH code base, there will always be at least $2^d$ of these chunks per kernel, as we loop over the cell itself as well as its vertex-connected neighbours. 
As we store connectivity information within the vertices, it is $2^d$. 
Otherwise, it would be $3^d$ for all neighbouring cells.
For adaptive meshes or setups where particles have massively differing search radii and hence have to be stored on multiple levels within the underlying spacetree, more chunks of cell data might have to be compared, i.e.~this is a lower bound on the chunks.

\subsection{Symmetry considerations}

Binary SPH kernels characteristically are symmetric or anti-symmetric.
For example, we have $F(p_1,p_2) = -F(p_2,p_1)$ for the forces between two particles.
Exploiting this symmetry reduces the overall compute cost per particle by a factor of two.
However, such symmetry is difficult to exploit along domain boundaries of a distributed memory parallisation, where we work with ghost particles mirroring information from another domain partition stored somewhere else.
It also requires careful design in the context of adaptive mesh refinement.

We refrain from explicitly using any symmetry in our code and henc avoid any additional algorithmic logic or bookkeeping on this level.
Instead, we exploit the vanilla $\mathcal{O}(N^2)$ code arrangement sacrificing the factor of $\frac{1}{2}$.

\subsection{Loop reordering}

Within each $\mathcal{O}(N^2)$ kernel, we can either loop over the active particles or the local particles first.
If the outer loop runs over the active particles, we fix one particle around the cell and compute its impact on all local particles of the kernel, i.e.~all particles within the cell of interest.
If the outer loop runs over the local particles, it runs over all particles within the cell and computes, per particle, its impact on all other praticles within the cell plus the surrounding area.

The active-local version accumulates the impact of forces or density contributions onto one local particle at a time.
The local-active version scattered the impact of one particle on its surrounding area.
Active-local yields very condensed write accesses yet introduces a reduction requiring synchronisation.
Local-active yields scattered writes.

\subsection{Multilevel, long-range interactions}

If we study long-range forces we have particles that do not interact exclusively with particles in neighbouring cells but also with particles far away.
We can introduce virtual particles, i.e.~multipoles \cite{White:1994:FMM}, to represent a cluster of local particles and hence reduce how many particles interact with particles far away, but we cannot eliminate the long-range data access of compute kernels.
While this challenge is out-of-scope here, similar data access patterns arise along adaptive mesh boundaries within the spacetree or in situations where particles have massively differing search radii.
In these cases, we store particles on different mesh resolution levels within the spacetree and hence introduce non-local fine-to-coarse data accesses.
To facilitate a parallel, task-based processing of the kernels, accesses to coarser levels are protected by atomics and semaphores.

We note that these multiscale interactions are relatively rare, i.e.~the bulk of compute workload stems from particle-particle interactions on the same level which yield very localised data access patterns.
It is hence possible to split the loops over particles \replaced[id=R2]{artificially}{artifcially} into two segments:
A preamble identifying which particle-particle interactions are local and which are non-local and hence require synchronisation primitives.
For very few particles, they represent long-range interactions.
We can filter out those interactions within the compute kernel and postprocess these interactions afterwards in a separate loop.

\subsection{\added[id=Ours]{Neutral elements vs.~branching}}

\added[id=Ours]{
 At the core of particle-particle interactions are distance checks:
 Particles have smoothing radii, and the two case distinctions \texttt{DensityPredicate} (Listing~\ref{algorithm:demonstrator:blueprint}) and its cousin \texttt{ForcePredicate} determine if particles are sufficiently close to each other to exchange information.
 Bucketing particles into cells adds an additional, conservative check on top of this.
 If two particles are not located within neighbouring cells, the predicate cannot hold. 
 We hence do not have to evaluate it.
}

\added[id=Ours]{
 Within the compute routines, we can skip the predicate evaluation as well. 
 All particle-particle interactions are subject to some scaling depending on the distance between two particles, and each kernel always sums up the contributions from all particle pairs.
 If we design the scaling to deteriorate to zero whenever the predicate does not hold, we can loop over all particle-particle pairs (within neighbouring cells):
 Particle-particle contributions over two particles that are too far away from each other become zero, i.e.~the neutral element, and will not contribute to the overall sum.
}

\added[id=Ours]{
 We eliminate branching. 
 In return we evaluate more interactions out of which many are eventually scaled with zero and hence make no contribution.
}

\section{Realisation of the code transformations through a compiler modification}
\label{section:realisation}

\added[id=copy]{
We implement our proposed techniques as a Clang LLVM front-end prototype, such that we can highlight their potential impact for the SPH demonstrator.
Clang takes C++ code and emits LLVM IR (intermediate representation).
This generated IR code still benefits from all LLVM-internal subsequent optimisation passes.
}

\added[id=copy]{
Clang's lexer, parser and semantic analyser yield an abstract syntax tree (AST).
The AST then is subsequently consumed to produce LLVM IR.
These steps constitute the compiler's \texttt{FrontendAction}.
We realise our functionality with a new \texttt{FrontendAction}.
}

\subsection{Data flow}

\added[id=copy]{
Our \texttt{FrontendAction} traverses the AST top down and searches for our annotations.
When it encounters a convert, it inserts allocation statements for the out-of-place memory allocations, it issues the actual data copying from (\ref{equations:AoStoSoA}), modifies the loop body $f$ into $\hat f$ according to (\ref{equation:view}), and eventually adds an epilogue that synchronises the data and frees the temporary buffers.
}

We need to identify all read and write accesses to struct members subject to the conversion.
Without major modifications to LLVM IR, these modifications are language-dependent.
Therefore, it is reasonable to collate all annotation handling exclusively within the \texttt{FrontendAction} rather than to scatter it among multiple compile stages.

Our automatic identification of views requires us to read whole code blocks before we can actually construct a tailored view representation.
Despite C++'s strict typing, information on the logical data types resulting from the narrowing is not yet available when we encounter a loop with annotations for the first time.
The notion of an automatic view identification violates the principle that the complete data information has to be at hand prior to the first usage.
Therefore, we need a preparatory pass through the source code to collect access data before we pass through the source code once again to apply all required modifications.
We have to analyse the AST before we start to enact any transformations.
The arising two-pass realisation of the front-end resembles preprocessor macros whose handling is realised as separate stage in the front-end, transparent to other components within the front-end pipeline.

By giving up on a single-read paradigm, i.e.~by running through the code multiple times, it becomes a natural choice to lower the annotated code onto plain OpenMP-augmented source code rather than to mangle GPU offloading into the translation process.
We apply a source-to-source translation followed by an additional OpenMP processing realised within the native, original Clang \texttt{FrontendAction}.

Despite being a logically separate pass through the AST, our \texttt{FrontendAction}'s mapping onto native, OpenMP-annotated C++ code can be realised as in-memory replacements of the source code.
This accommodates the fact that the LLVM AST itself is almost immutable.
\added[id=copy]{
We modify the in-memory buffers that hold the original files that make up the translation unit after they are parsed once, using the additional annotations to guide all rewrites.
Subsequently, the compiler re-parses these in-memory buffers, and feeds that new AST into the compiler pipeline.
This realisation reuses the preprocessor, lexer, parser and semantic analysis in both passes.
}

\added[id=copy]{
Within our intermediate code generation, the names of all temporary variables are mangled, using the original container identifiers, to avoid variable re-declarations.
For the actual loop body, we then redirect the memory accesses to $\mathbb{A} \cup \mathbb{\hat A}$ to an intentionally non-mangled helper variable that masks the original AoS data accesses.
}

\added[id=copy]{
As we precede Clang's original \texttt{FrontendAction}, it is possible to dump
our output into source files explicitly instead of implicitly passing it on into
the subsequent \texttt{FrontendAction}.
While originally designed for debugging, this feature is particularly appealing
in environments where the modified compiler is only available locally,
while the compilers on the target production platform cannot be
modified.
}

\added[id=copy]{
As we navigate the source code using the AST, the effect of our annotations is scoped at the translation-unit level:
The data transformations cannot propagate into user translation units (object files) and notably fail to propagate into libraries.
}

\subsection{A minimally invasive data structure transformation}

As we realise all data transformations within the compiler, it is an option to let the compiler rewrite the functions' loop bodies explicitly:
We parse the body $f$ and generate a modified $\hat f$ over the view.
This modified loop body then is subject to further optimisation passes.
This strategy is efficant and useful \cite{Radtke:2024:AoS2SoA}, but it can involve significant code rewrites, making is difficult to maintain.
A further disadvantage of a genuine rewrite is that it becomes more difficult to debug and profile the resulting code, as it might become unclear which code parts correspond to which original source code instruction. 
This is a common property of any precompiler and code-generation approach. 


For our present compiler realisation, we propose to use proxy objects~\cite{Gamma:1994:DesignPatterns}.
In this implementation, a proxy object preserves the structure and method signatures of the original AoS type, i.e.~it includes all static methods, type aliases, typedefs, using declarations, and constexpr values from the original type, as well as provides the storage and member functions required to maintain syntactic backwards-compatibility with the original AoS data structures. Yet, the proxy object replaces all value-holding struct members (fields) with corresponding references to locations within the SoA buffers.
Since static struct members and methods do not affect the memory layout, we employ a blanket approach, i.e.~make them plain forwards to the proxied class.

Once the proxy object type is available, we can instantiate a proxy object at the start of the loop body using the original loop ``counter'' (iterator's) variable name.
This shadows all accesses to the underlying data and instead redirects them through the proxy.
It is a very simple transformation boiling down to one command inserted into $\hat f$.
The remainder of the loop body remains syntactically invariant.
The prologue now creates the SoA data representation, i.e.~the view, while the proxy object redirects data accesses via references into the out-of-place converted data rather than the original container for all \replaced[id=R2]{data members}{data mambers} from $\mathbb{A}_{\text{in}} \cup \mathbb{A}_{\text{out}}$. The remaining data members and non-static methods are omitted from the proxy object entirely.

LLVM’s optimisation passes eliminate the indirect data accesses through the proxy object, generating optimised assembly equivalent to that produced by whole-loop rewriting.
From hereon, LLVM can also identify vectorisation potential.
The overall approach is a lightweight code-generation rather than a heavy compiler-based rewrite of major code parts (cmp.~alternative proposed in \cite{Radtke:2024:AoS2SoA}).
Indeed, we delegate the heavy lifting behind code optimisation to subsequent compiler passes and realise a strict separation-of-concern that frees the developer from the cumbersome and error-prone manual creation and management of proxy objects.

\begin{lstlisting}[language=C++,
                   label=algorithm:rewrite-demo:vanilla,
                   caption=Element-wise addition of two attributes over an AoS loop.,
                   basicstyle=\footnotesize]
struct Data { 
  double a, b, _unused;
};

void kernel(Data *buf, int size) {
  [[clang::soa_conversion_target("buf")]]
  for (int i = 0; i < size; i++) {
    buf[i].a += buf[i].b;
  }
}
\end{lstlisting}

The two-stage realisation of this concept---attribute realisation from Section~\ref{section:annotations} and code generation---hosts the complicated logic in the first source code pass, while the second pass is a rather mechnical code rewrite and source code generation (cmp.~Listing~\ref{algorithm:rewrite-demo:vanilla} against Listing~\ref{algorithm:rewrite-demo:rewritten}). 
The first analysis pass has to identify all fields and methods of the AoS type accessed within a kernel. 
For this, all source code employed within a loop kernel has to be visible (cmp.~notes in Section~\ref{section:annotations}).
As many scientific codes, our SPH benchmark is heavily templated-based.
For template-based kernels, \replaced[id=R1]{the}{th} view analysis is deferred until the compiler instantiates the template. 
A recursive, queue-driven algorithm then tracks all method calls originating from the kernel, enqueuing methods where the structure escapes and detecting cycles to avoid infinite recursion. 



\algrenewcommand\algorithmicindent{0.4em}%

\begin{lstlisting}[language=C++,
                   label=algorithm:rewrite-demo:rewritten,
                   caption=Transformed loop from Listing~\ref{algorithm:rewrite-demo:vanilla}.,
                   basicstyle=\footnotesize]
void kernel(Data *buf, int size) {
  // define the SoA proxy view 
  struct Data_view_3360 {
    double &a;
    double &b;
    Data_view_3360 &operator[](int i) { return *this; }
    Data_view_3360 &operator*() { return *this; }
  };

  // helper data structure to handle the [] operator
  struct Data_SoaHelper_3360 {
    double *a;
    double *b;
    Data_view_3360 operator[](int i) {
      return Data_view_3360{a[i],b[i],};
    }
  };

  // allocate the SoA buffer
  alignas (64) char __soa_buf_3360[16 * size];
  double * __restrict__ a_3360 
    = (double*) (__soa_buf_3360 + 8 * size);
  double * __restrict__ b_3360 
    = (double*) (__soa_buf_3360 + 0 * size);

  // prologue - populate SoA buffers
  for (int i = 0; i < size; i++) {
    a_3360[i] = *((double*) (((char*) &buf[i]) + 0));
    b_3360[i] = *((double*) (((char*) &buf[i]) + 8));
  }

  // original loop
  for (int i = 0; i < size; i++) {
    // helper variables
    // buf masks accesses to AoS data
    auto Data_SoaHelper_instance_3360
      = Data_SoaHelper_3360{a_3360,b_3360,};
    auto buf = Data_SoaHelper_instance_3360[i];
  
    buf[i].a += buf[i].b;
  }

  // epilogue - populate AoS data
  for (int i = 0; i < size; i++) {
    *((double*) (((char*) &buf[i]) + 0)) = a_3360[i];
  }
}
\end{lstlisting}


\subsection{Data management}

Out-of-place transformations for containers typically require heap data allocations if the containers' size is unknown at compile time.
Heap allocations can become expensive if issued very frequently in high-concurrency codes \cite{Hager:2011:HPCIntro}.
However, our prologue-epilogue paradigm does not require strictly dynamic data structures.
We know the size of the underlying container by the time we hit the prologue while the struct member set $\mathbb{A}_{\text{in}} \cup \mathbb{A}_{\text{out}}$ is a compile-time property.
Since the size of the container does not change while we traverse the loop, we work with variable-length arrays (VLAs)~\cite{Cheng:1995:VariableLengthArrays}. This approach lets us avoid making memory allocator calls entirely (cmp.~Listing~\ref{algorithm:rewrite-demo:rewritten}).

\subsection{OpenMP}

The transformation pass itself does not modify the loop beyond the standard SoA transformation.
If offloading annotations are used, it decorates these transformed loops with the appropriate OpenMP target offloading pragmas.
It leverages standard OpenMP. 
This is possible as we keep all data conversions on the host, work with an out-of-place data transformation, and make the generation of plain C++ code follow a complete view analysis.
The compiler ``knows'' exactly how much data in which format is held in the temporary views' data structures, and that all of these data are held in plain SoA.

Due to the ownership over the temporary data, it is possible to deploy the responsibility for all correct data movements to the OpenMP subsystem.
The compiler then has to generate offloading-capable binaries, and it is up to the user to correctly configure the compilation environment. 
This includes setting up the necessary paths and environment variables specified by hardware vendors for both compilation and runtime execution.
Our solution remains hardware-agnostic.

\subsection{GPU data handling}

Issuing device memory allocation commands on a critical path can lead to substantial performance degradataion~\cite{Wille:2023:GPU} which necessitates implementing a memory management scheme that safely reuses accelerator allocations. Our implementation lazily allocates per-thread, per-transformed-loop buffers that grow depending on the sizes of the working sets. To minimise synchronisation overheads, we schedule all accelerator-related operations, including memory transfers and kernel executions asynchronously using the standard OpenMP \texttt{nowait depend(...)} constructs. Since the prologue-loop-epilogue triad is a logically synchronous operation, we may not introduce asynchronicity that is observable beyond the immediate context of the transformed loop. This necessitates at least one host-device synchronisation for every invocation of the loop which we implement using the standard OpenMP \texttt{taskwait} construct. The synchronisation, together with the per-thread, per-loop buffer allocation guarantees data safety.

\subsection{Interplay with other translation passes and shortcomings}

By confining our transformation entirely within Clang, we ensure that all optimisations performed by the LLVM optimiser as part of subsequent passes are applied to the transformed kernels. 
At this point, our transformations are already out of the way. 
This guarantees that each kernel is processed in a single, consistent form, eliminating potential conflicts arising from multiple interpretations of the same code:
If functions are used within views and then also without views, we logically have already replicated the functions' realisation and the optimisations handle the replica independently.

Such a design not only avoids unfortunate side-effects.
It notably ensures that HPC-critical optimisations, such as auto-vectorisation, are applied independently and in a bespoke, tailored way to the vectorisation-friendly SoA representations.

A limitation of our approach is its dependence on complete visibility of all functions that interact with the transformed AoS data. 
We need access to the whole loop body's call graph to identify all write and read accesses to AoS data members and to be able to produce a modified loop kernel.
Making the whole call hierarchy visible becomes challenging when external functions, either directly or indirectly referenced by the kernel, reside outside the analysed translation unit. 
In such cases, our transformations break down.

A potential remedy involves leveraging link-time optimisation (LTO) to extend the transformation’s reach beyond a single translation unit. However, this approach introduces several practical challenges. 
Firstly, it is only viable for external functions with accessible source code.
Precompiled library functions remain beyond reach. 
Secondly, full LTO (fat LTO) is required to provide a comprehensive view of all compiled code, whereas many large codebases favour thin LTO due to its superior compilation speed.

\section{Results}
\label{section:results}

We assess the impact of our compiler prototype on two architectures.
Our first system is an Intel Xeon Platinum 8480+ (Sapphire Rapids) testbed. 
It features $2 \times 56$ cores over $2 \times 1$ NUMA domains spread over
two sockets, hosts
an L2 cache of 2,048 KByte per core and a shared L3 cache with 105 MByte per
socket.
Our second testbed is an Nvidia GH200 (Grace Hopper) system.
It features 72 cores in a single socket configuration.
Each core has an L2 cache of 1,024 KByte, and a shared L3 cache with 114 MByte for the entire
socket. The system contains the Nvidia H200 GPGPU chip boasting 96GB HBM3 memory at 4TB/s.

\begin{table}[htb]
  \caption{
    Overview of benchmarked SPH kernels. We report on the total contribution to the runtime, the cell-local compute complexity, and the size of the active sets.
    \label{table:results:kernel-overview}
  }
  \begin{center}
    \begin{tabular}{l|rlrrrr}
Name & Runtime & Compl. & \phantom{xx} $|\mathbb{A}_{\text{in}}|$ \phantom{xx} & \phantom{xx} $|\mathbb{A}_{\text{in}}|_{\text{byte}}$ \phantom{xx} & \phantom{xx} $|\mathbb{A}_{\text{out}}|$ \phantom{xx} & \phantom{xx} $|\mathbb{A}_{\text{out}}|_{\text{byte}}$ \\
  \hline
  Density &
  	39.0\% & 
  	$\mathcal{O}(N^2)$ &
  	9 & 88 &
  	6 & 48 \\
  Drift &
  	2.2\% & 
  	$\mathcal{O}(N)$ &
  	3 & 20 &
  	2 & 12 \\
  Force &
  	58.5\% & 
  	$\mathcal{O}(N^2)$ &
  	13 & 128 &
  	4 & 40 \\
  First kick &
  	0.1\% & 
  	$\mathcal{O}(N)$ &
  	4 & 48 &
  	3 & 32 \\
  Second kick &
  	0.2\% & 
  	$\mathcal{O}(N)$ &
  	11 & 112 &
  	8 & 80 \\
  \hline
\end{tabular}
  \end{center}
\end{table}

We benchmark the core SPH algorithm and split up all measurements into data for the individual compute kernels that dominate the runtime (Table~\ref{table:results:kernel-overview}).
Meshing overhead, sorting, I/O and other effects are excluded.
\replaced[id=Ours]{Our}{In line with previous discussions, our} SPH code does not refine the underlying spacetree down to the finest admissible level~(cmp.~related design decisions in \cite{Schaller:2024:Swift}).
Instead, we impose a cut-off:
If the number of particles per cell ($ppc$) underruns a certain threshold, \replaced[id=Ours]{we stop decomposing the mesh further}{the underlying mesh is not decomposed further}.
This way, codes can balance between meshing overhead and algorithmic complexity.
In our result, we benchmark setups with average $ppc \in \{64,128,256,512,1024\}$.
Further to that, we assess configurations where the particles are scattered over the heap in main memory against configurations where the particles are stored in chunks of continuous memory and hence allow for coalesced memory access.

\added[id=Ours]{
 Our spacetree code employs a mixture of domain decomposition and task-based parallelism.
 Consequently, the compute kernels per cell are called in parallel by all threads, and
 we do not need to exploit multiple cores within a kernel anymore.
 However, we should be able to exploit one core's vector capabilities or offload each kernel onto a GPU. 
}

Throughout the presentation, all data are normalised as cost per particle update, which might comprise interactions with all cell-local neighbours (density, force) or simply denote the runtime divided by the loop count.
All data are always employing all cores of \replaced[id=Ours]{one socket, i.e.~we eliminate potential NUMA effects.}{the compute node, i.e.~the underlying code base is parallelised.
The data are hence node benchmarks or employ the whole GPU respectively.}

\subsection{Baseline code}

%
%
We start our studies with a simple assessment of the baseline kernel realisations without our compiler-guided transformations.
It is important to obtain a sound performance baseline for fair comparisons.
Besides this, it is not clear \deleted[id=Ours]{in the context of data format transformations} \replaced[id=R1]{whether}{if} certain realisation patterns are particularly advantageous for the data re-organisation or challenge them.
Therefore, we distinguish various different kernel variants:
\begin{itemize}
  \item In the scattered generic kernel, we stick to the vanilla blueprint of Algorithm~\ref{section:demonstrator} and let the particles be scattered all accross the heap. Each cell hosts a sequence of pointers to addresses all over the place.
  \item A second variant sticks to the generic loop over C++ containers yet ensures that the particles of one cell are stored continuously in memory. \deleted[id=Ours]{However, this knowledge is not exploited explicitly.}
  \item Finally, we use this knowledge about continuous chunks and split up each loop into a loop over chunks. Per chunk, we then employ a plain \deleted[id=R1]{parallel for} \added[id=R1]{loop} with pointer arithmetics, i.e.~we expose the fact that coalesced memory access is possible explicitly within the source code.
  \item For the $\mathcal{O}(N^2)$ kernels, we distinguish two kernel variants of the latter approach: In the variant labelled \texttt{local-active}, we first run over the local particles in the outer loop before we traverse the active particles. In the variant with the label \texttt{active-local}, we permute these two outer loops.
  \item \added[id=Ours]{For the $\mathcal{O}(N^2)$ kernels, we furthermore have realisations with branching (e.g.~see \texttt{DensityPredicate} in Listing~\ref{algorithm:demonstrator:blueprint})) available as well as a realisation without branching where the predicates effectively yield masking.}
\end{itemize}

\begin{figure}[htb]
  \begin{center}
    \includegraphics[width=0.6\textwidth]{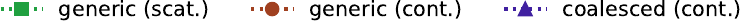}
    \includegraphics[width=0.45\textwidth]{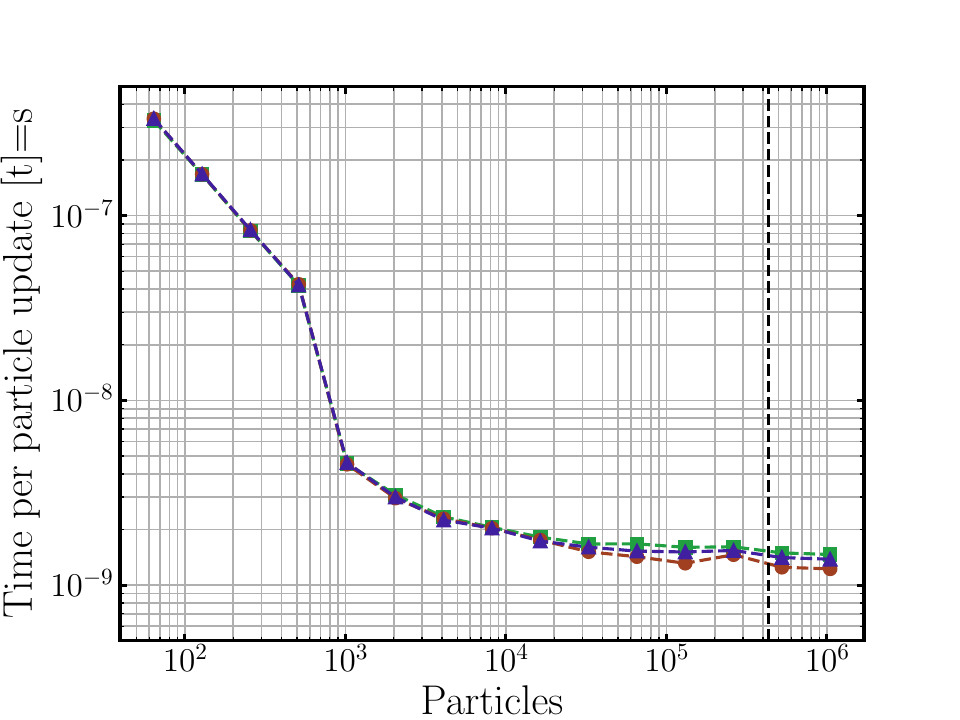}
    \includegraphics[width=0.45\textwidth]{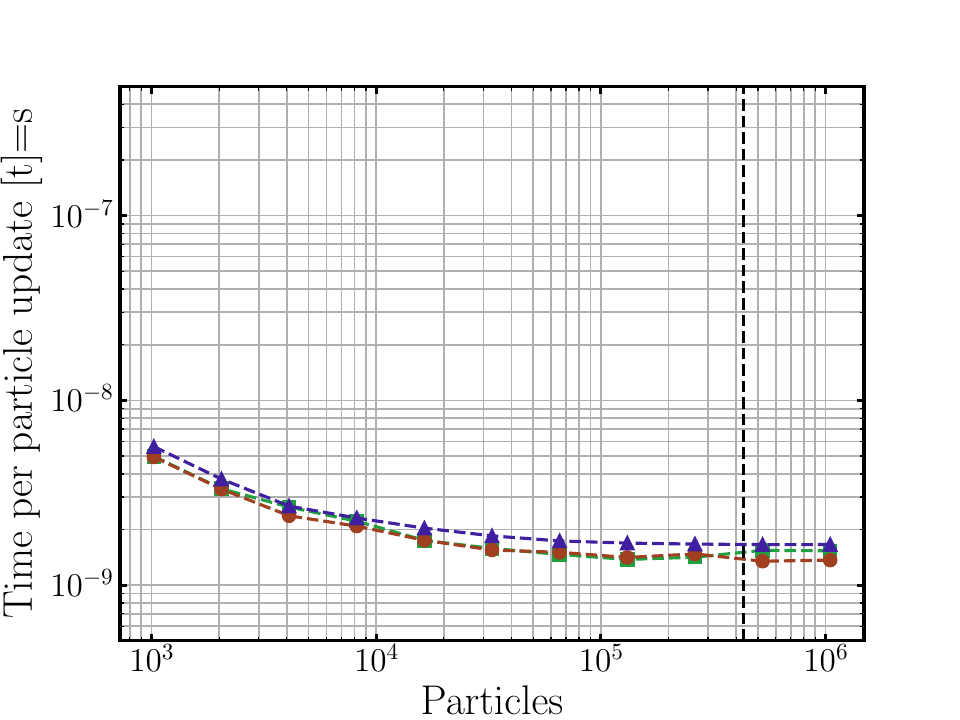}
    \includegraphics[width=0.45\textwidth]{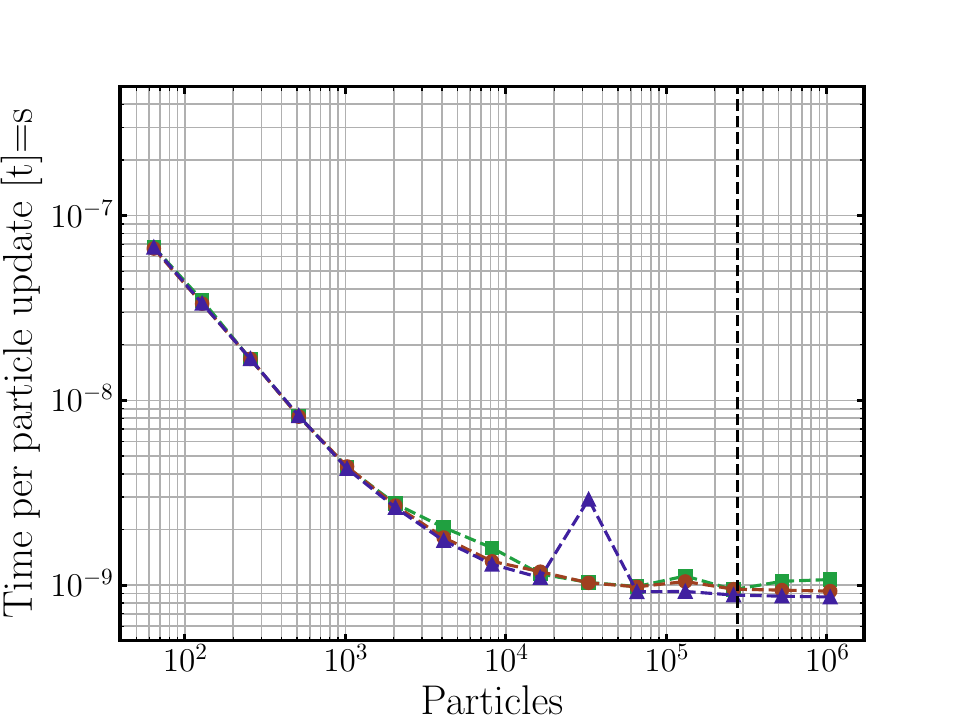}
    \includegraphics[width=0.45\textwidth]{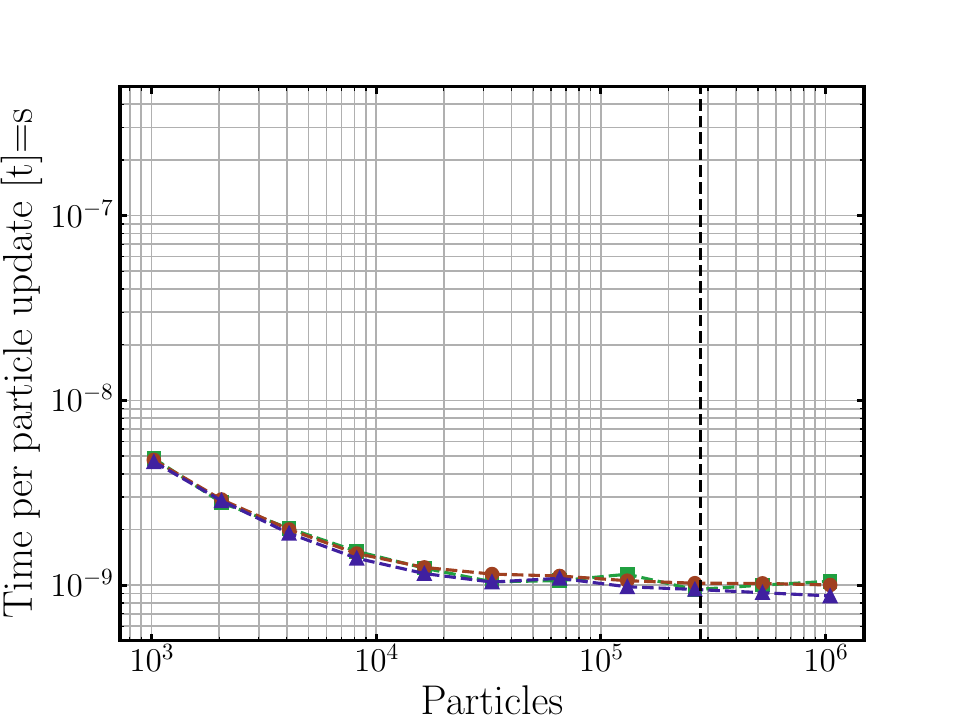}
  \end{center}
  \caption{
    \added[id=R2]{
   	Cost per particle update for the drift kernel on the Intel (top) and Grace (bottom) CPU node without any data transformations. 
   	We distinguish $ppc=64$ (left) from $ppc=1024$ (right). The vertical line denotes \replaced[id=R1]{how many particles the accumulated L2 caches could hold}{the sum of all L2 caches}.
   	}
   \label{figure:results:baseline-drift-implementation-variants-with-ifs}
  }
\end{figure}

%
%
\begin{figure}[htb]
  \begin{center}
    \includegraphics[width=0.99\textwidth]{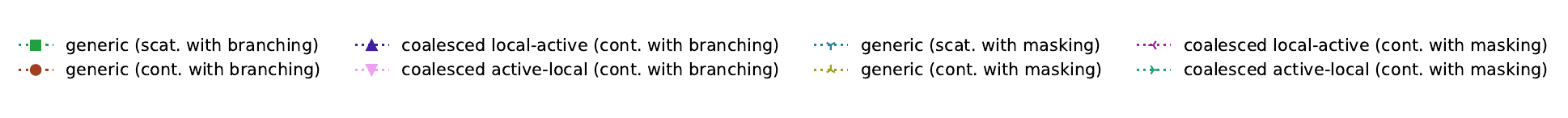}
    \\[-0.2cm]
    \includegraphics[width=0.45\textwidth]{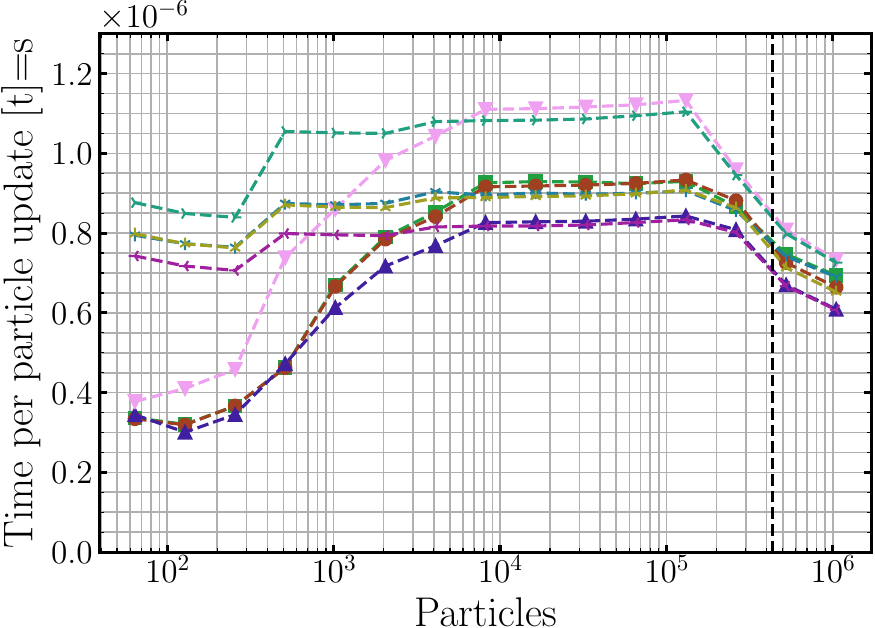}
    \includegraphics[width=0.41\textwidth]{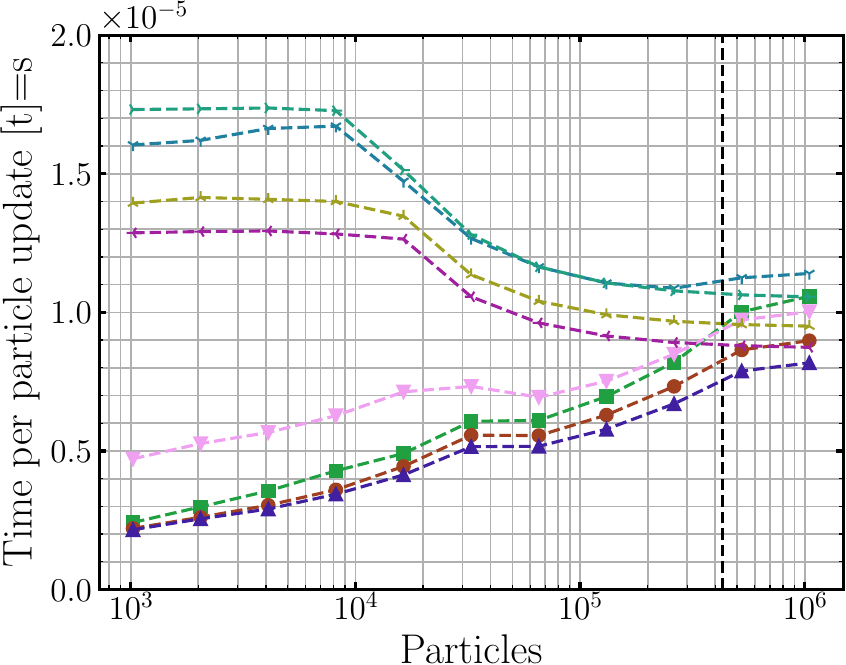}
    \includegraphics[width=0.45\textwidth]{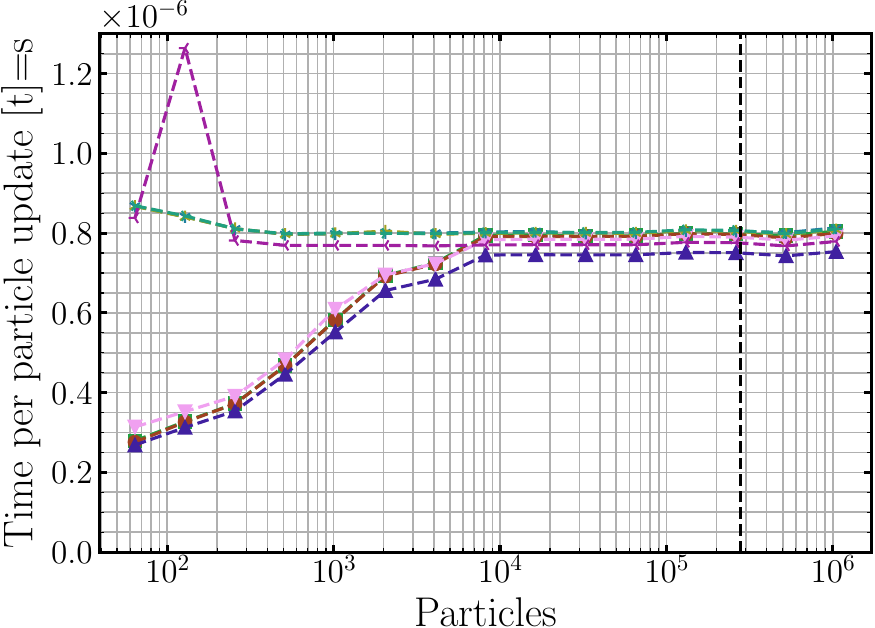}
    \includegraphics[width=0.41\textwidth]{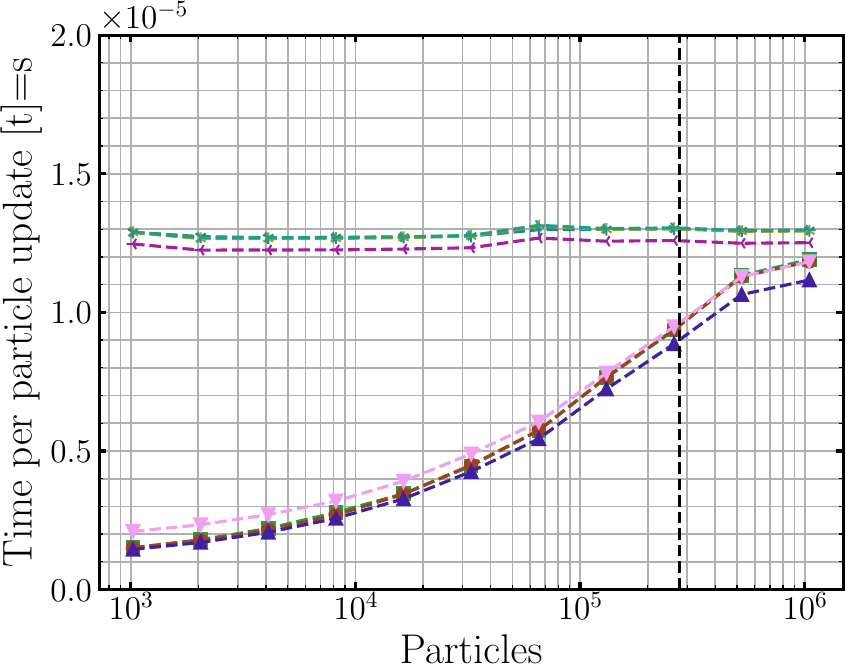}
  \end{center}
  \caption{
   	Cost per particle update for \replaced[id=R2]{the density}{various} kernels on the Intel (top) and Grace (bottom) CPU node without any data transformations. 
   	We distinguish $ppc=64$ (left) from $ppc=1024$ (right).
   \label{figure:results:baseline-density-implementation-variants}
  }
\end{figure}

%
%
\replaced[id=Ours]{
For a plain implementation of the linear compute kernels, 
}{
For all setups,} 
the throughput improves with increasing number of particles (Figure~\ref{figure:results:baseline-drift-implementation-variants-with-ifs})\deleted[id=Ours]{ unless we have a very small total particle count}.
The \added[id=R2]{arrangement of the memory and, hence, memory access pattern (scattered vs.~coalesced) as well as the two} different kernel realisation variants make no significant difference.

\replaced[id=Ours]{The}{while the} kernels with local $\mathcal{O}(N^2)$ complexity are significantly more expensive than their linear counterparts \added[id=Ours]{(Figure~\ref{figure:results:baseline-density-implementation-variants})}.
\replaced[id=Ours]{The}{Notably, the} fact that data are available as continuous chunks makes \replaced[id=R2]{a minor}{no major} difference\added[id=Ours]{ on Intel and no observable difference on ARM, unless we explicitly exploit the coalesced memory (Listing~\ref{algorithm:kernel-variants:coalesced}).}
 \deleted[id=R1]{no matter if we try to exploit this fact or not}
\added[id=Ours]{
 Replacing the branching over predicates with masking, i.e.~multiplications by zero, is not a advantageous, as long as we can work within the close-by caches.
 Once the problems become significantly large and do not fit into the local L2 caches anymore, both the implementation with case distinctions and masking yield comparable performance. 
 For all variants, it is better to loop over the local particles receiving density contributions first, and then to let the inner loop accumulate all the density contributions.
 Analogous observations can be made for the force calculation.
}
\deleted[id=Ours]{
The ARM CPU yields slightly noisier data than the x86 counterpart.
A small workload per kernel improves the throughput of kernels with quadratic complexity, but larger $ppc$ counts favour the linear kernels.
Falling out of the L2 caches has no performance impact and hence cannot explain any runtime behaviour.
}

%
%
The plateauing for larger particle counts \added[id=Ours]{for the linear kernels} suggests that there is some compute management overhead which is amortised as the overall workload increases.
We pick the minimal total particle count such that all CPU cores are always busy, i.e.~there are always enough kernel invocations to employ all threads.
However, workload imbalances between the threads can better be smoothed out with a larger total set of compute kernels.
The overhead stems merely from a load distribution argument.
The local $\mathcal{O}(N^2)$ character of the force \added[id=R2]{(not shown)} and density calculation explains why they benefit from small $ppc$ (we effectively have a complexity of $\mathcal{O}(ppc^2)$), while the kicks \added[id=R2]{(not shown)} and drifts are streaming kernels and hence do benefit from larger loops which pipe data through the system.
\added[id=Ours]{
 For the quadratic kernels, the cost per particle increases as we increase the total particle count, as the cells do not to fit into one local cache anymore and we stress the lower memory levels.
 We eventually plateau as we become solely L3-bound.
 The decrease of cost for large particle counts and small $ppc$ on the Intel system is something we cannot explain.
}

As the kernels with quadratic local complexity dominate the runtime (Table~\ref{table:results:kernel-overview}), \replaced[id=Ours]{most simulation codes favour}{the simulation code benefits from} small $ppc$.
\deleted[id=Ours]{
We assume that noise in the measurements and notably cost peaks stem from situations where we fall out of caches, or setups where a lot of the guard predicates that determine \replaced[id=R1]{whether}{if} a particle is to be updated return false.
}
\added[id=Ours]{
 Vectorisation seems to make no major difference.
}
The \replaced[id=Ours]{limited}{lack of} impact of continuous data seems to be surprising, but is reasonable once we take into account that our baseline data structure is AoS. 
The particles might be continuous in memory, but each kernel picks only few data members $\mathbb{A}_{\text{in}}$ to read and very few $\mathbb{A}_{\text{out}}$ to write (Table~\ref{table:results:kernel-overview}).
Consequently, the data access pattern is scattered even though the underlying data are continuous.
There is no coalesced memory access.

\begin{observation}
 A better utilisation of the hardware is inferior to a reduction of the local compute complexity of the kernels\deleted[id=Ours]{ as long as there are enough of these kernels in flight}. 
\end{observation}

%
%

\noindent
The classic HPC metric MFlop\added[id=Ours]{s}/s suggests to use reasonably large $ppc$ to balance for the throughput between the different phases.
Once we take the characteristic distribution of the compute phases into account and weight the improvements accordingly, it becomes clear that, from a science per time point of view, we have to work with as small $ppc$ as possible, which implies using aggressive \replaced[id=R1]{adaptive mesh refinement (AMR)}{AMR} in our case\replaced[id=Ours]{. An analogous argument holds for the predicates which should be evaluated as early as possible to avoid obsolete calculations}{;}
even though this challenges the kernel \added[id=Ours]{vector} efficiency.

\subsection{Data transformation impact\added[id=R2]{ on generic kernel implementations}}

%
%
We next enable the data compiler-guided transformations\replaced[id=R2]{.
 Where appropriate, we present normalised speedups, i.e.~how the performance changes due to the compiler.
 All timings always comprise all conversion overhead, i.e.~are end-to-end measurements.
}{
  and compare the outcomes to the CPU baseline.
  This allows us to quantify the transformation overhead.
}

\begin{figure}[htb]
  \begin{center}
    \includegraphics[width=0.6\textwidth]{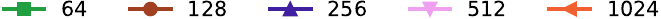}
    \includegraphics[width=0.45\textwidth]{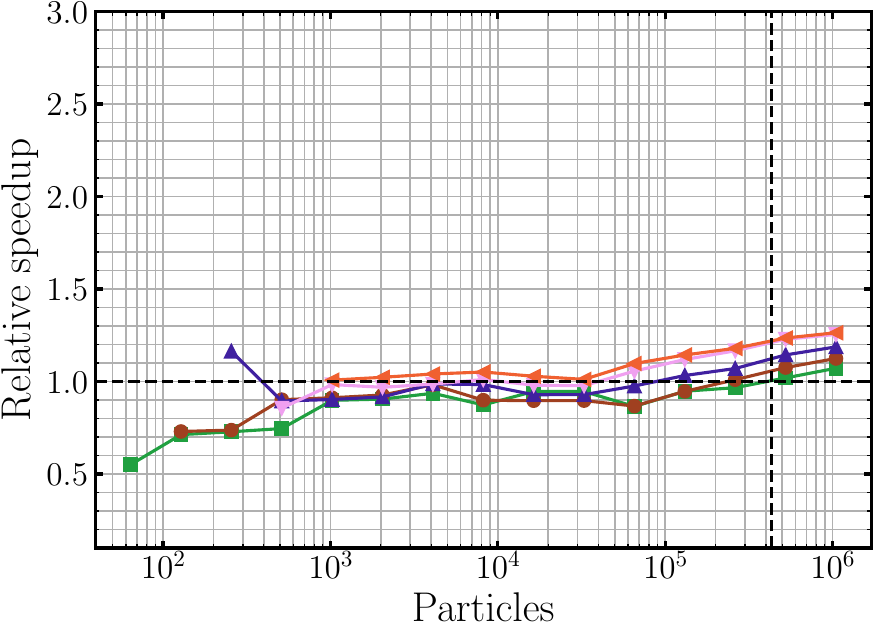}
    \includegraphics[width=0.42\textwidth]{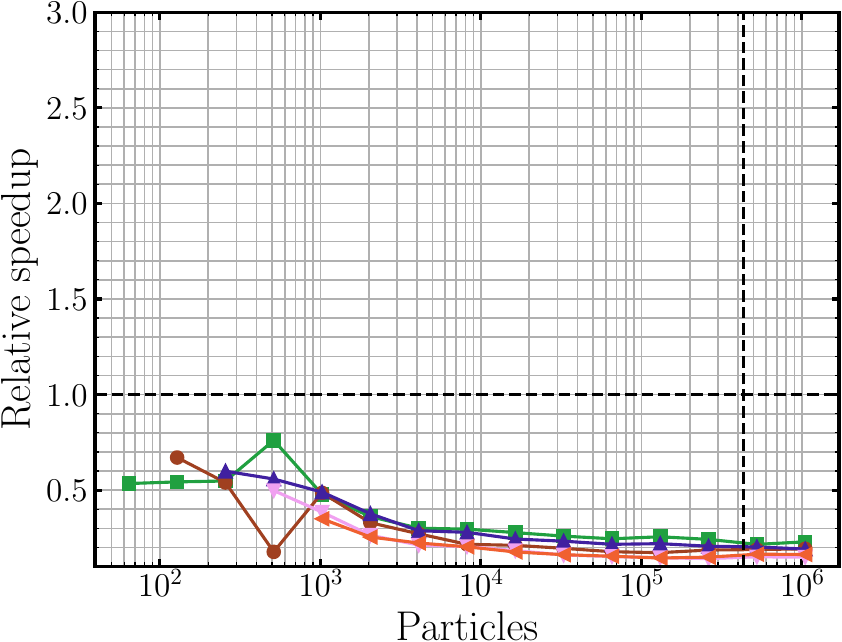}
    \includegraphics[width=0.45\textwidth]{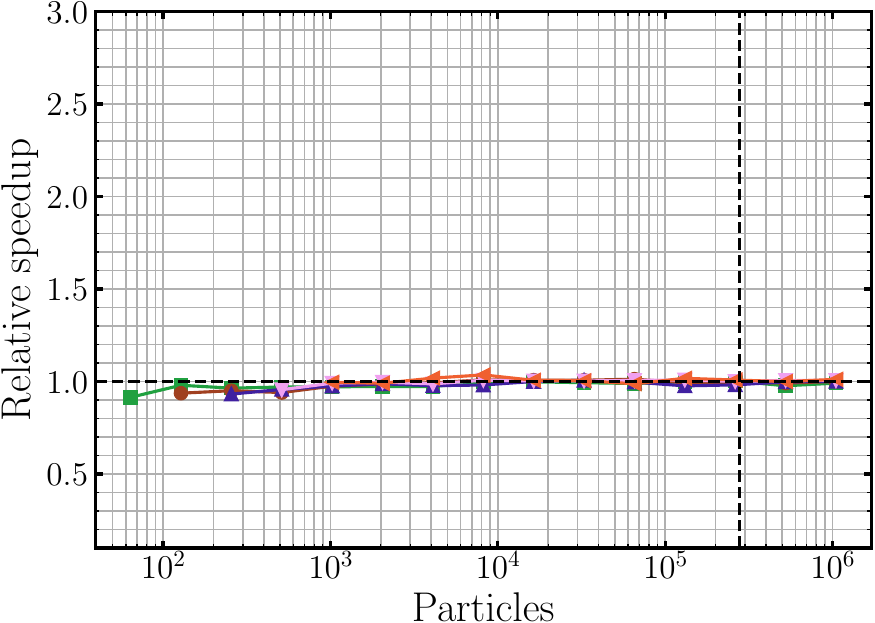}
    \includegraphics[width=0.42\textwidth]{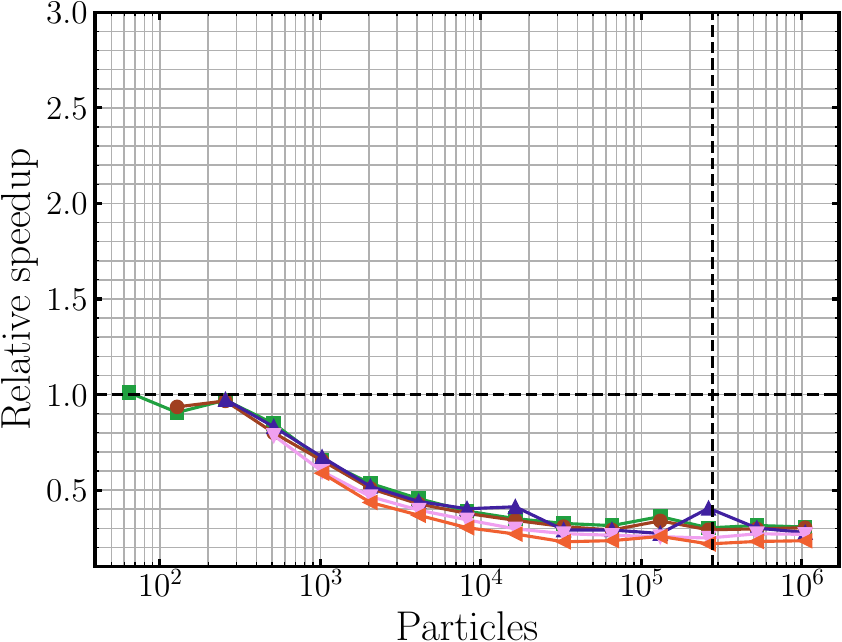}
  \end{center}
  \caption{
    \added[id=R2]{
     Relative change of performance due to the compiler transformations for the Intel (top) and Grace (bottom) testbed.
     We present data for the density kernel (left) and the drift kernel (right) for various particles per cell ($ppc$).
     All measurements employ a generic implementation over the scattered data.
    }
   \label{figure:results:transformation-impact-scattered}
  }
\end{figure}

%
%
\deleted[id=Ours]{The introduction of the compiler transformations smoothes out the measurements and leads to very stable plateaus (Figure~\ref{figure:results:transformation-impact-scattered}).}
For the kernels with linear compute complexity, \replaced[id=R2]{our compiler transformations}{they} introduce an overhead which can result in a performance degradation \replaced[id=R2]{
of up to 75\% if we run the kernels over scattered memory (Figure~\ref{figure:results:transformation-impact-scattered}). For small problems, this
}{
of up to an order of magnitude on the x86 chip. This
}
overhead penalty is less pronounced on the ARM chip.
For the quadratic kernels, we gain an \replaced[id=R2]{moderate runtime improvement of up to 25\%}{order of magnitude in speed} on the Intel platform\added[id=R2]{ if we host a sufficiently large number of particles in each cell}, while the ARM chip \replaced[id=R2]{shows no speedup}{yields even more significant speed improvements}.

\begin{figure}[htb]
  \begin{center}
    \includegraphics[width=0.6\textwidth]{experiments/kernel-throughput-2/gi001/cell_legend.pdf}
    \includegraphics[width=0.45\textwidth]{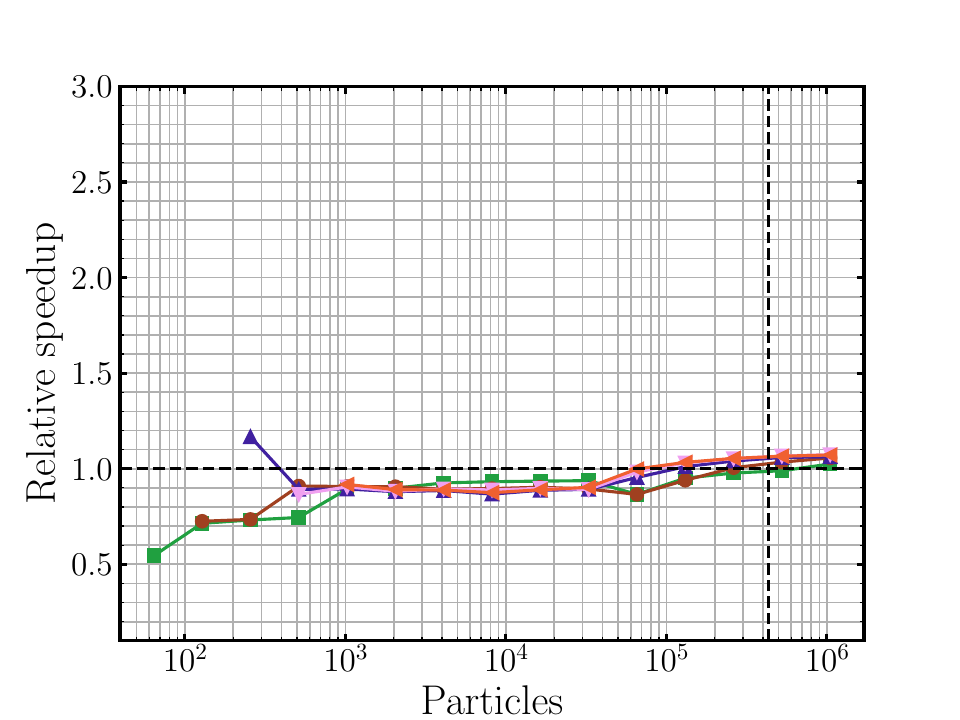}
    \includegraphics[width=0.45\textwidth]{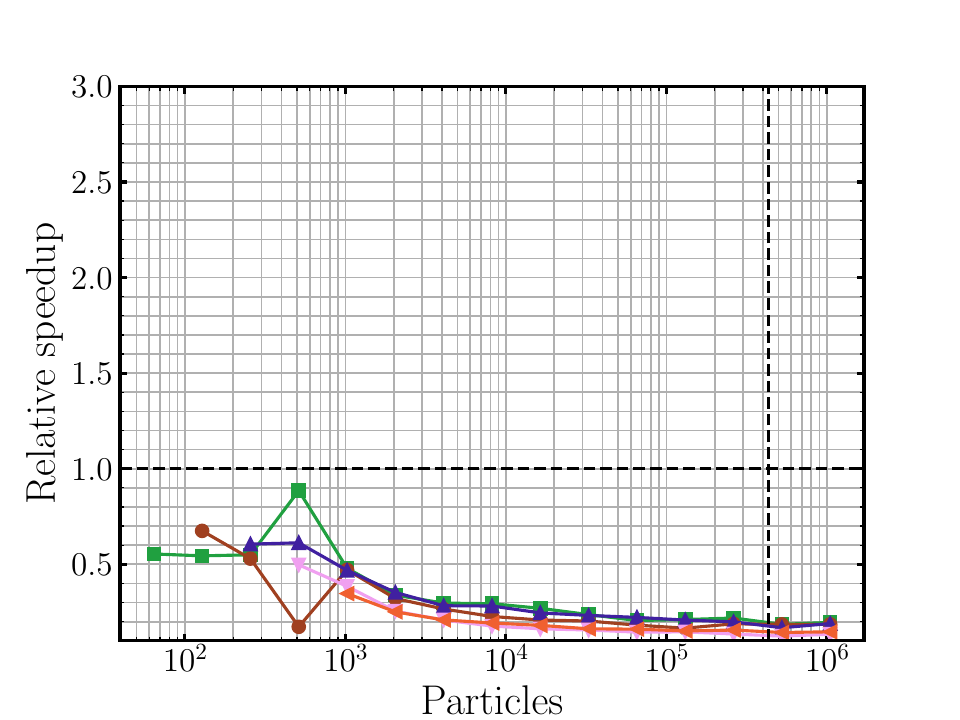}
    \includegraphics[width=0.45\textwidth]{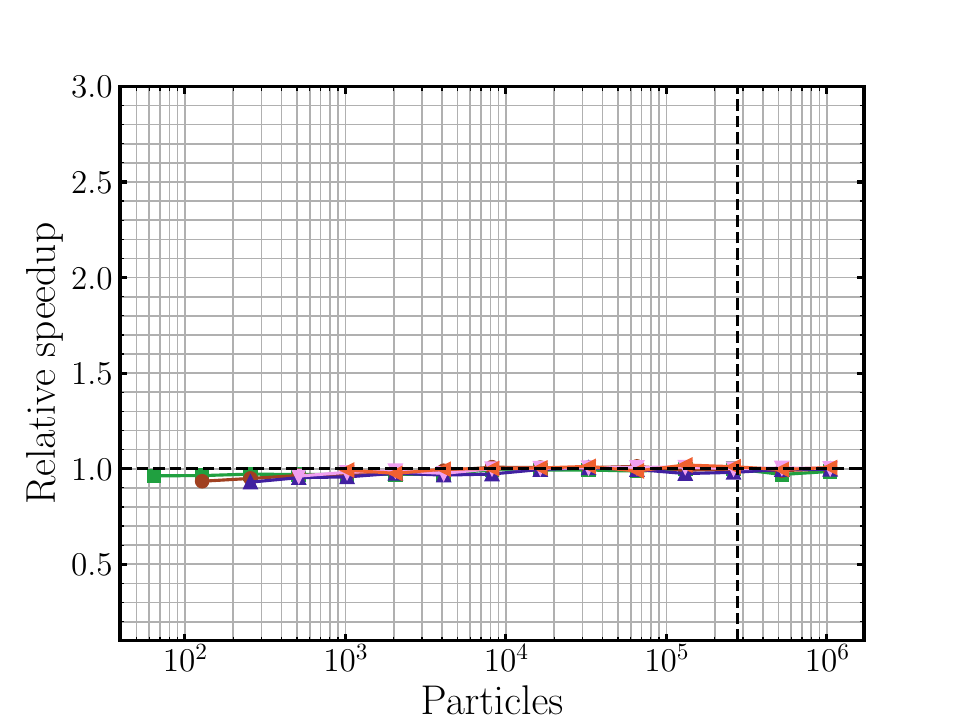}
    \includegraphics[width=0.45\textwidth]{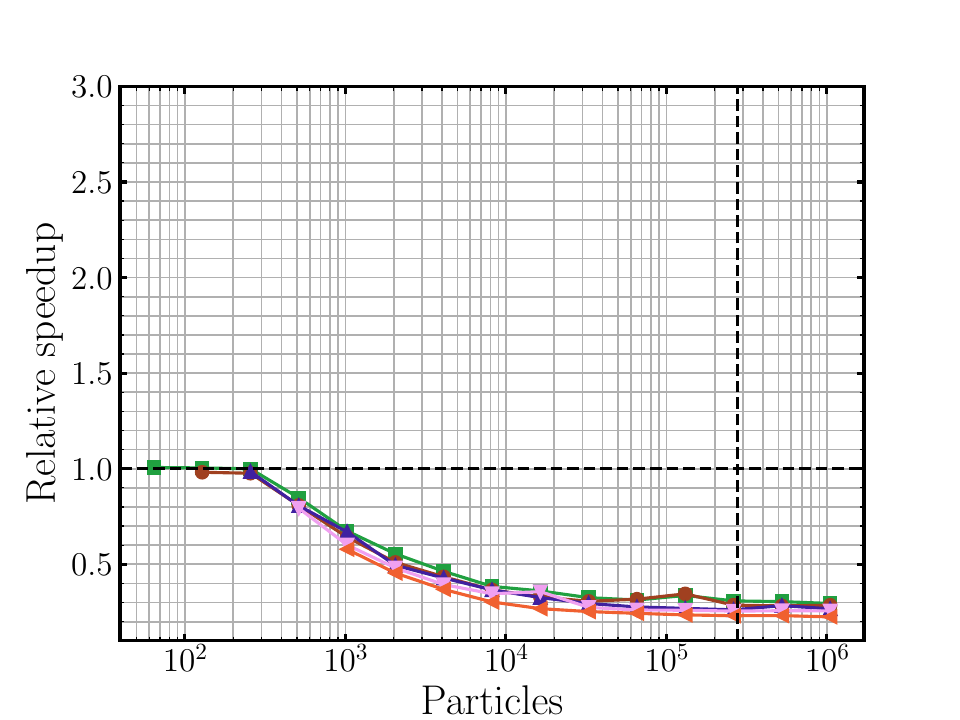}
  \end{center}
  \caption{
    \added[id=R2]{
     Relative change of performance due to the compiler transformations for the Intel (top) and Grace (bottom) testbed.
     We present data for the density kernel (left) and the drift kernel (right) for various particles per cell ($ppc$).
     Different to Figure~\ref{figure:results:transformation-impact-scattered}, we employ the generic kernels over continuous data, i.e.~the memory access is coalesced even though this knowledge is not explicitly exploited in the source code (Listing~\ref{algorithm:kernel-variants:coalesced}).
    }
   \label{figure:results:transformation-impact-coalesced}
  }
\end{figure}

%
%
Having containers over coalesced memory \replaced[id=R2]{makes the transformation yield close to no difference to the runtime anymore for the quadratic compute kernels (Figure~\ref{figure:results:transformation-impact-coalesced})}{seems to make no big difference, but trying to exploit this fact makes the throughput deteriorate.
Once we zoom into different kernel variants and plot their relative speedup over the vanilla baseline version without any views (Figure~\ref{figure:transformation-overhead:normalised-speedups}), it becomes clear that the coalesced memory organistion does help.
It just is not relevant for all $ppc$ and kernel choices, i.e.~notably makes a difference for smaller $ppc$}.
Cache blocking is implicitly triggered by our out-of-place reordering.
We therefore eliminate a lot of memory transfer noise\replaced[id=Ours]{
 for the scattered memory arrangement.
 If the data are continuous in memory right from the start, this performance gain evaporates.
 The other way round, the transformations compensate for scattered data layouts.
}{
. As
each kernel picks only few attributes from the baseline container, our prologues and epilogues all continue to yield scattered memory access and hence fail to benefit from a continuous arrangement of the data in memory.
However, they do benefit from very large, uninterrupted loops:
}

On both architectures, the linear kernels benefit from larger $ppc$, as they can stream data through the chip.
\replaced[id=Ours]{
 Sole streaming however implies that we cannot benefit from temporary data reordering. 
}{
The same effect materialises for the quadratic kernel once we enable our transformations:
If we reduce $ppc$ as we split the loop iteration ranges into chunks, we reduce the efficiency of the transformations.
}

The ARM chip is slow \deleted[id=Ours]{and weak} compared to its Intel cousin yet \replaced[id=R1]{comes}{come up} with a better balanced memory bandwidth relative to the compute capability.
Hence, the \replaced[id=R1]{conversions}{conversion} have \replaced[id=Ours]{close to no positive impact for the quadratic kernels. 
The architecture can cope well with scattered memory access.}{a smaller impact}.
\deleted[id=Ours]{Overall, we obtain impressive speedups exceeding one order of magnitude.}

\begin{observation}
 Continuous ordering of the \added[id=Ours]{AoS} input does not consistently have a major positive impact on the runtime once our annotations are used. 
 \deleted[id=Ours]{In return, the split into tiny subchunks trying to exploit continuous data access is counterproductive, i.e.~the arising loop blocking makes the performance deterioriate.}
\end{observation}

%
%
\noindent
\deleted[id=Ours]{Loop blocking is a classic performance engineering strategy, notably once combined with a continuous arrangement of the underlying memory.}
Our data suggests that \replaced[id=Ours]{the empiric lesson learned to always arrange data continuously has}{these empiric lessons learned have} to be rethought or even turned around as we introduce compiler-based temporary reordering with views\added[id=Ours]{ over AoS data}.
\deleted[id=Ours]{Loop ordering is disadvantageous.}
Sorting data \replaced[id=Ours]{might even be}{is} disadvantageous\added[id=Ours]{ as soon as we acknowledge that it is typically not free. It is better to do this resorting locally and temporarily}.
However, it is not clear \replaced[id=R1]{whether}{if} we can uphold this observation for all particle types, i.e.~number of struct members, deeper memory hierarchies or other algorithms.
\deleted[id=Ours]{Notably SPH algorithm flavours with a lot of long-range interactions might behave different if the ``long-range data'' falls out of L3 cache.}

\added[id=R1]{
 The percentage of time spent on memory layout transformation varies by kernel (Table \ref{table:results:runtime-breakdown}). Average statistics computed over all variants demonstrate that our linear kick and drift kernels experience a runtime overhead comparable to the cost of the computation itself. The same is not true for the pair-wise force and density kernels where the computational cost dominates.
}

\begin{table}[htb]
  \caption{
    Relative time spent on conversion, computation and back-conversion per kernel after our compiler transformations have been applied. 
    \label{table:results:runtime-breakdown}
  }
  \begin{center}
    \begin{tabular}{l|rrrr}
Name & AoS $\rightarrow $ SoA & \phantom{xx} Computations & \phantom{xx} SoA $ \rightarrow $ AoS \\
  \hline
  Density &
  	1.80\% & 
  	97.5\% &
  	0.70\% \\
  Drift &
  	47.8\% & 
  	26.1\% &
  	26.1\% \\
  Force &
  	1.22\% & 
  	98.4\% &
  	0.34\% \\
  First kick &
  	28.1\% & 
  	50.9\% &
  	21.0\% \\
  Second kick &
  	32.5\% & 
  	48.6\% &
  	18.9\% \\
  \hline
\end{tabular}

  \end{center}
\end{table}

\subsection{\replaced[id=Ours]{Data transformations on kernels over continuous memory}{Kernel modifications on CPUs}}

\begin{figure}[htb]
  \begin{center}
    \includegraphics[width=0.6\textwidth]{experiments/kernel-throughput-2/gi001/cell_legend.pdf}
    \includegraphics[width=0.45\textwidth]{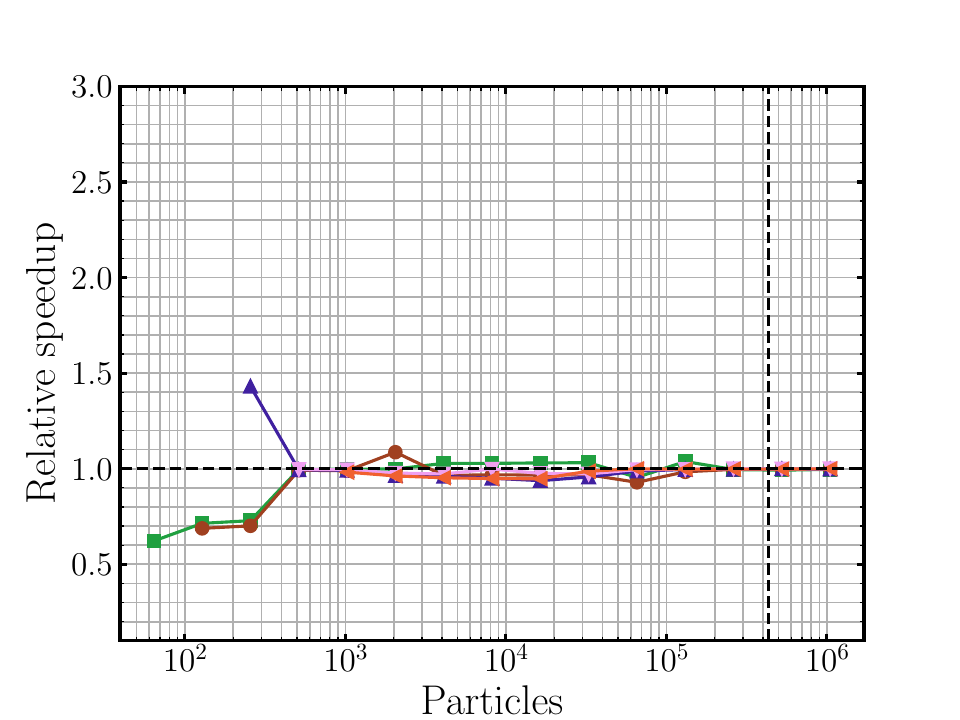}
    \includegraphics[width=0.45\textwidth]{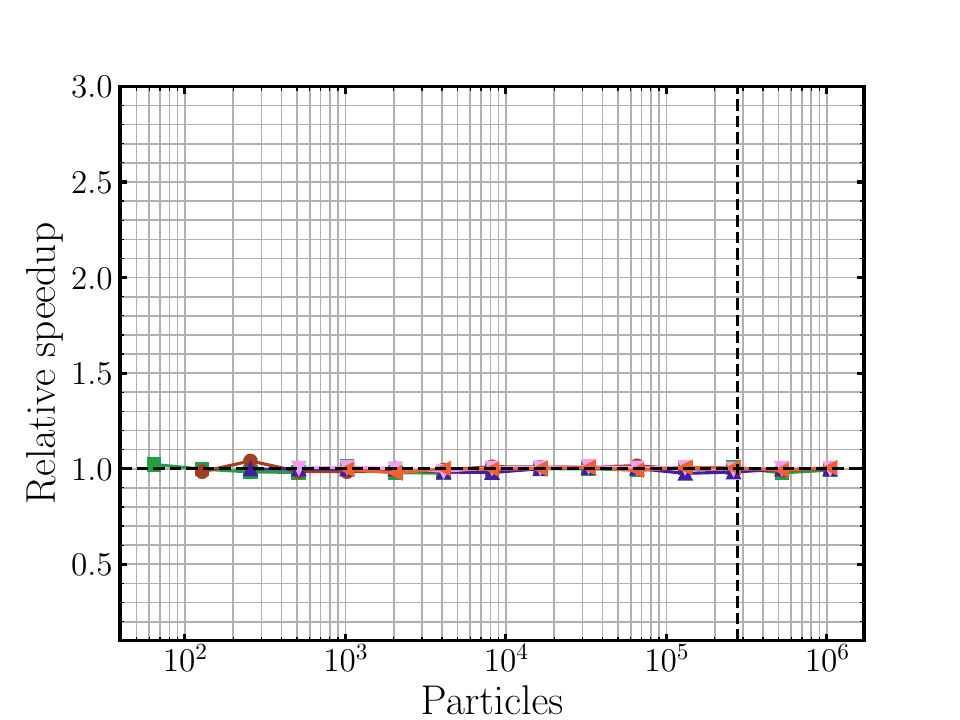}
  \end{center}
  \caption{
    \added[id=R2]{
     Relative change of performance of the density kernel due to the compiler transformations on the Intel (left) and Grace (right) testbed,
     if the kernel implementation exploits the fact that the underlying data per cell are ordered continuously in memory.
    }
   \label{figure:results:cpu-kernel-coalesced}
  }
\end{figure}

\added[id=Ours]{
 The vanilla versions of our kernel implementations (Listing \ref{algorithm:demonstrator:blueprint}) has no internal knowledge on consecutive memory arrangements. 
  Explicitly exploiting coalesced memory (Listing~\ref{algorithm:kernel-variants:coalesced}) gives us a faster baseline code (Figure~\ref{figure:results:baseline-drift-implementation-variants-with-ifs}).
  However, it also implies that the data transformation impact becomes negligible (Figure~\ref{figure:results:cpu-kernel-coalesced}).
}

\begin{observation}
 \replaced[id=Ours]{The}{In return, the} split into tiny subchunks trying to exploit continuous data access is counterproductive, i.e.~the arising loop blocking makes the performance \added[id=Ours]{improvements due to the compiler transformations} deterioriate.
\end{observation}

\noindent
\added[id=Ours]{
 Once AoS is already arranged (pre-sorted) to facilitate a stream-like memory access pattern, it is not clear anymore whether it is advantageous to work with coalesced memory directly or to let the compiler transformations switch to SoA with views.	
}

\subsection{Kernel modifications on CPUs}

\begin{figure}[htb]
  \begin{center}
    \includegraphics[width=0.6\textwidth]{experiments/kernel-throughput-2/gi001/cell_legend.pdf}
    \includegraphics[width=0.45\textwidth]{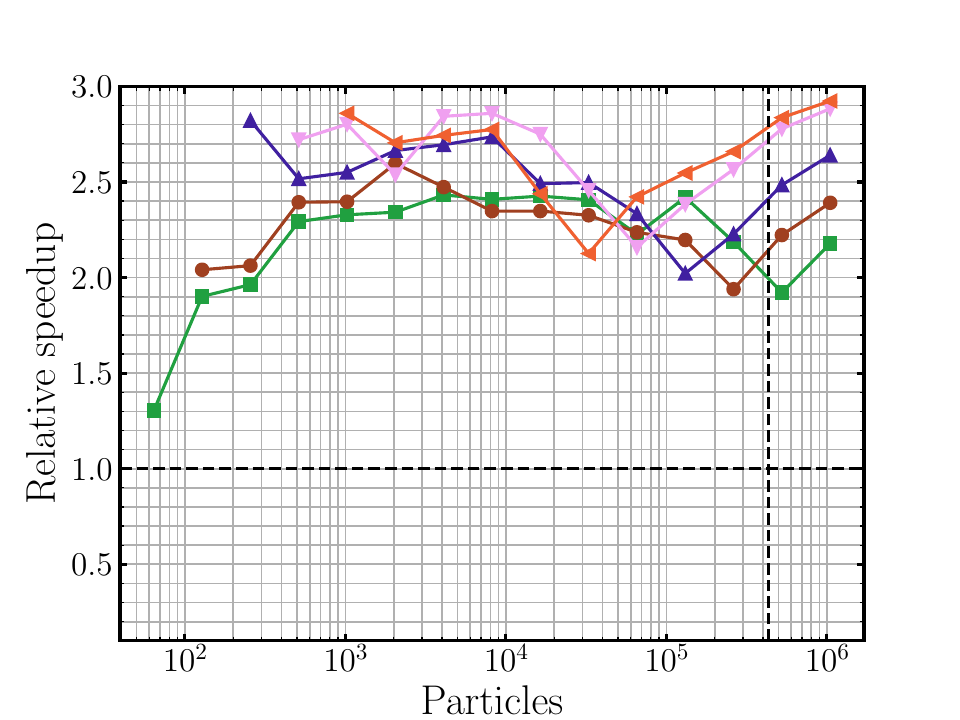}
    \includegraphics[width=0.45\textwidth]{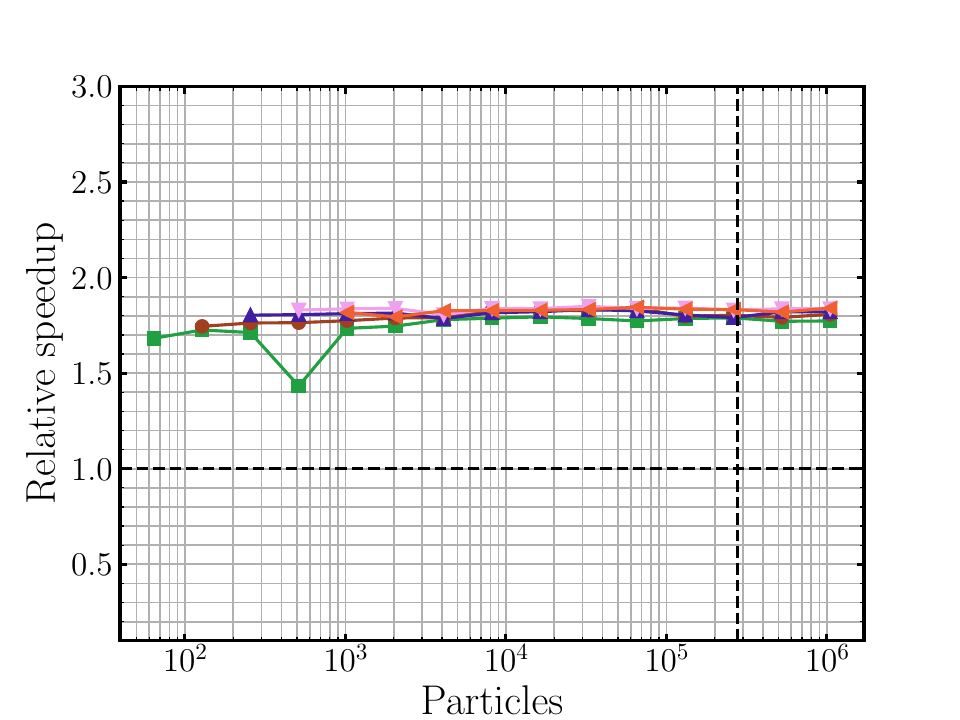}
  \end{center}
  \caption{
    \added[id=R2]{
     Speedup of the density kernels on Intel (left) and Grace (right), when we exploit coalesced memory, replace the branching with masking and apply our compiler transformations. 
    }
   \label{figure:results:cpu-kernel-scattered-vectorised}
  }
\end{figure}

\begin{figure}[htb]
  \begin{center}
    \includegraphics[width=0.6\textwidth]{experiments/kernel-throughput-2/gi001/cell_legend.pdf}
    \includegraphics[width=0.45\textwidth]{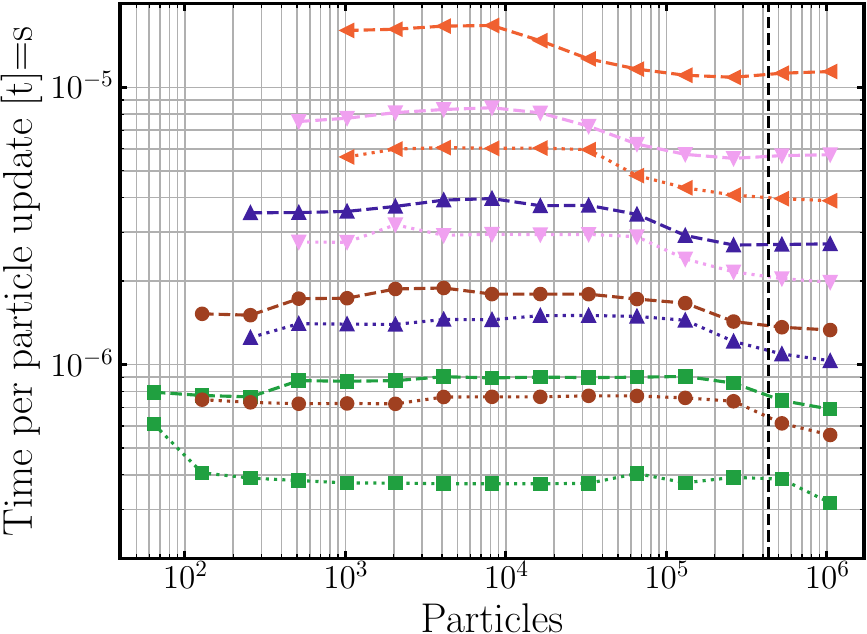}
    \includegraphics[width=0.45\textwidth]{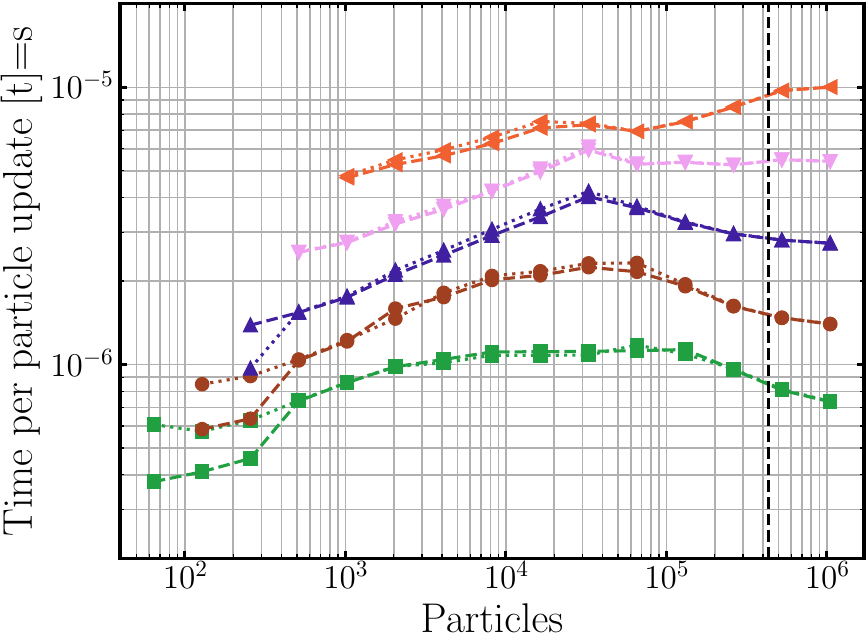}
    \includegraphics[width=0.45\textwidth]{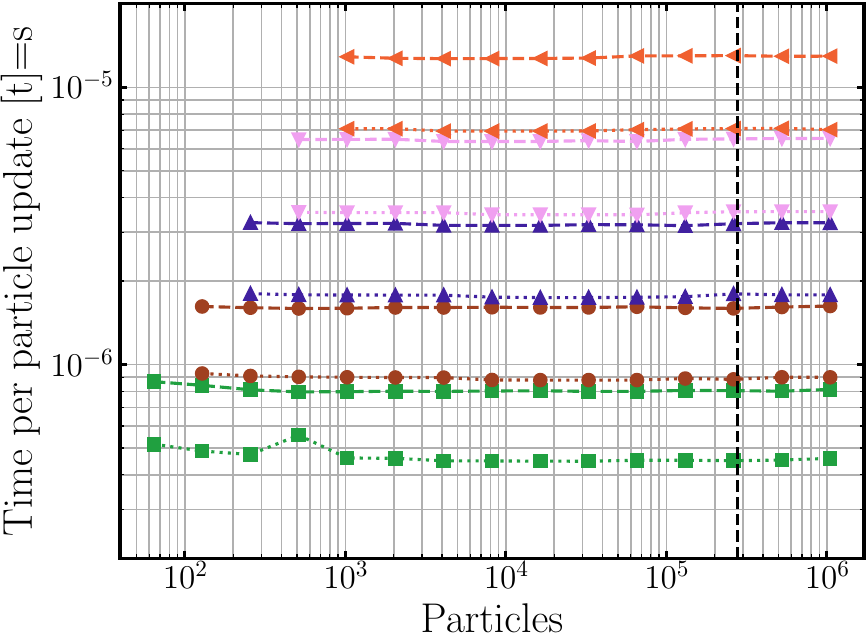}
    \includegraphics[width=0.45\textwidth]{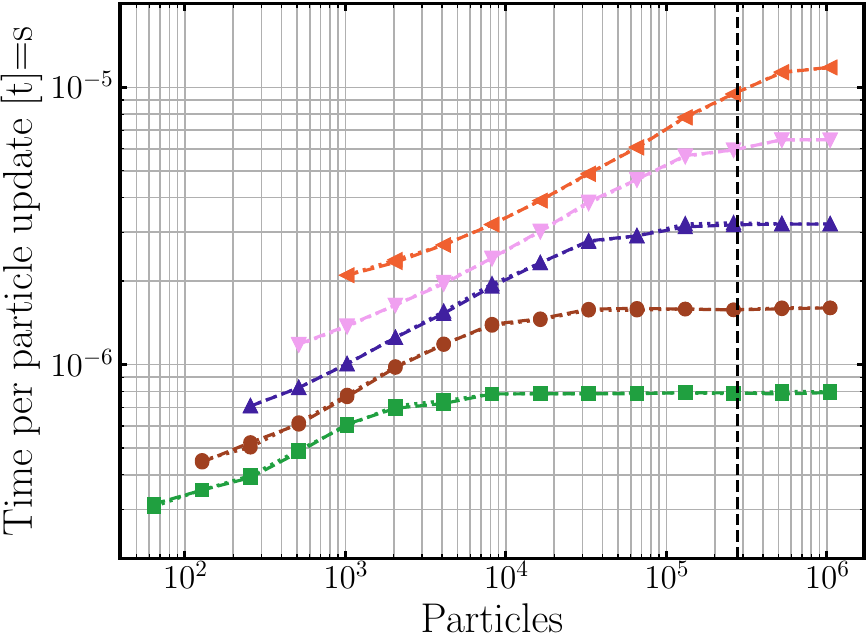}
  \end{center}
  \caption{
    \added[id=R2]{
     Absolute runtimes for various $ppc$ for the density kernel on Intel (top) and the Grace (bottom) test system.
     We compare the kernel performance over scattered data plus a kernel with masking (left) 
     against the fastest kernel variant without any compiler transformations over continuously ordered data (right).
     For all variants, we show the performance of a natively compiled version (dashed) and the version with compiler transformations (dotted).
    }
   \label{figure:results:cpu-total-comparison}
  }
\end{figure}

Our SPH compute kernels are heavy on comparisons yielding branching\deleted[id=Ours]{ and even involve some internal atomic operations}.
Therefore, we have to assume that they do not vectorise accross multiple particle--particle evaluations.
\deleted[id=Ours]{
It is a straightforward code modification to remove the handling of long-range interactions and instead only to evaluate a boolean within the loop that flags ``is there an interaction requiring an atomic access''.
If this flag is set, we can loop over the particle pairs again and evaluate the long-range interaction.
Our assumption is that most kernels do not enter any branch with an atomic.
Eliminating or masking out the neighbour predicate (such as \texttt{DensityPredicate} in Listing~\ref{algorithm:demonstrator:blueprint}) is more difficult.
We however can artificially remove the if statement and study the impact of such a tweak on the performance.
}
The \replaced[id=Ours]{replacement of branching with explicit masking}{removal of the atomic entries} yields a speedup of \replaced[id=R2]{two to three}{around a factor of two} on Intel systems (Figure~\ref{figure:results:cpu-kernel-scattered-vectorised}) for the $\mathcal{O}(N^2)$ kernels.
The ARM system showcases a \replaced[id=R2]{less pronounced}{similar} effect.

\deleted[id=R2]{Eliminating the (physically required) distance checks however does not give us a faster code.
The elimination of the atomics helps the local-active version over coalesced memory accesses to almost close the gap to the generic code base.
However, once we normalise all speedups (Figures~\ref{figure:cpu:overhead:Grace} and \ref{figure:cpu:overhead:Intel}), it becomes clear that a generic implementation not trying to exploit any memory insight performs best---although an advantageous memory layout can help (cmp.~differing observations in [\ldots]).
Yet, this statement is only true for the kernels with quadratic compute complexity. 
The linear kernels with a Stream-like access pattern do not benefit from the view concept~[\ldots].
}

\begin{observation}
 The SoA conversion with views helps the compiler to generate vector instructions. 
\end{observation}

%
%
\noindent 	
A speedup of a factor of two is not what we would hope to see on a modern vector architecture for a compute-intense kernel.
\deleted[id=Ours]{However, the SPH interactions are complex and still contain branching, and our implementation does not dive deeper into a kernel assessment and optimisation.
The data showcases that it is reasonable to remove complex atomic accesses from the core loops if these are infrequently encountered.
The data also highlights that it is important to avoid accummulation steps:
Fixing one particle and gathering the impact from all surrounding particles is not fast.
Instead, it is better to realise a spread-out paradigm, i.e.~scattered writes outperform accumulation once the data is converted into SoA.
However, small subchunks---with an average $ppc \approx 64$ and continuous particle chunks being assigned to the $2^d=8$ vertices\added[id=R1]{ per cell}, each chunk has only around eight particles---still hamper perforance and it remains advantageous to gather all data into a temporary large SoA memory block.
The removal (or masking out) of interactions improves the vector efficiency quite significantly.
Yet, it also increases the compute load, and eventually fails to compensate for the increased amount of work. 
}
\added[id=R2]{
 However, this improvement is sufficient to let the generic implementation over totally scattered data outperform our fastest manual implementation  relying on ordered, continuous AoS data (Figure~\ref{figure:results:cpu-total-comparison}).
}

%
%
\deleted[id=Ours]{While the views help the compiler to vectorise better, it is not a free lunch.
It is merely a first step to facilitate better vectorisation and to kickstart more in-depth kernel analysis how to improve the performance further. 
}

\begin{observation}
 A temporary, local conversion into SoA is beneficial for kernels with high computational complexity\replaced[id=Ours]{, as long as their implementation in principle facilitates very aggressive vectorisation}{ but problematic for streaming kernels}. 
\end{observation}

\noindent
\added[id=R2]{
 For the optimised quadratic kernels relying on sorting the AoS data such that all data access is coalesced, our transformations fail to improve the performance further.
 For an arbitrary kernel over scattered data, a realisation using masking instead of branching allows the transformations to yield highly vectorised code with coalesced memory access and good memory access behaviour which eventually outperforms the manually tuned kernels.
}
\subsection{GPU offloading}

We conclude our studies with offloading kernels to the GPU.
As our SPH kernels are \replaced[id=Ours]{pure functions without any side-effects}{well-defined and atomic}, this can be done through annotations only, i.e.~we do not have to employ additional OpenMP pragmas.
With the Grace-Hopper superchip, we have two variants available how to manage the data transfer:
We can copy over data explicitly via map clauses, or we can use the hardware's shared memory paradigm which translates into a lazy memory transfer similar to a cache miss.

\begin{figure}[htb]
\centering
    \includegraphics[width=0.6\textwidth]{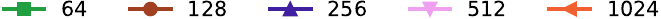}
    \includegraphics[width=0.45\textwidth]{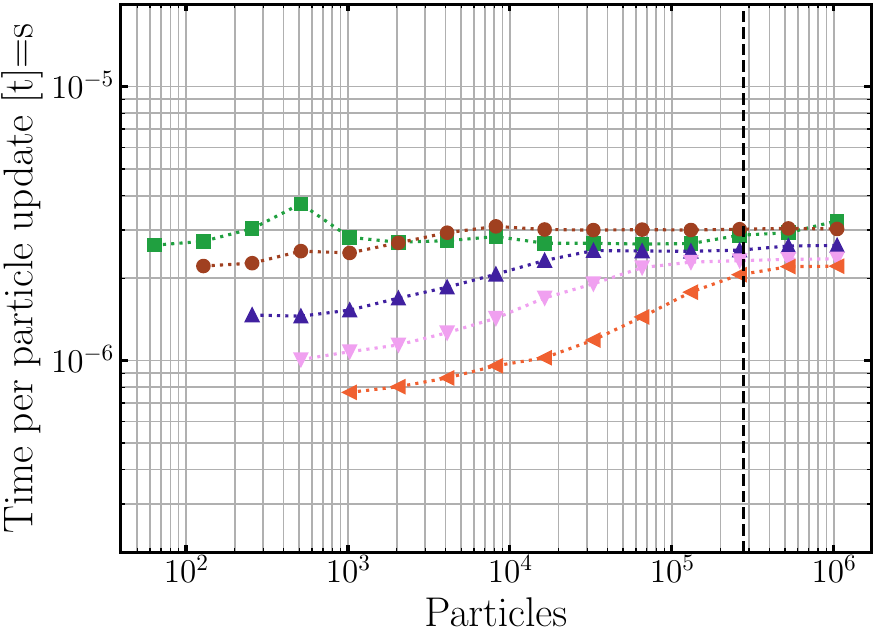}
    \includegraphics[width=0.45\textwidth]{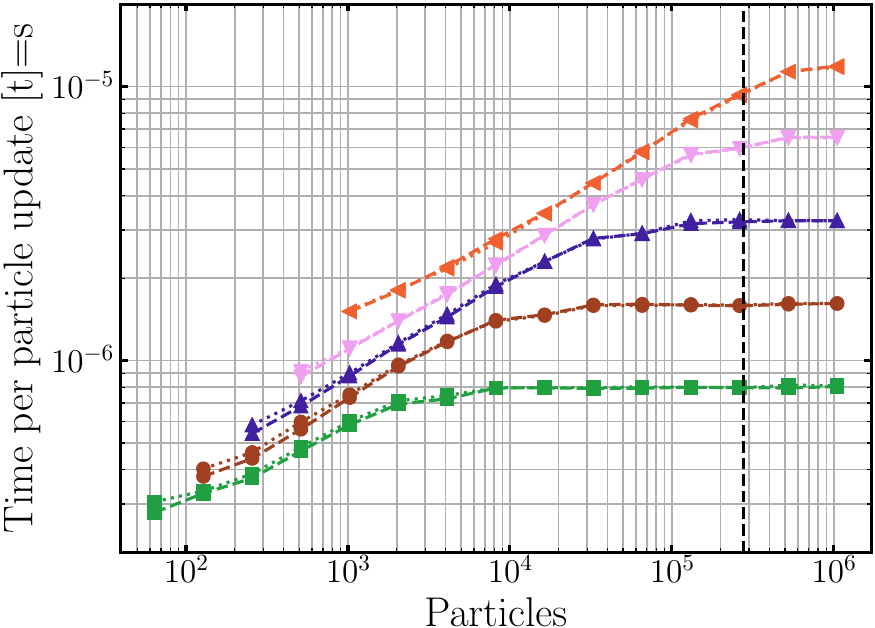}     
    \includegraphics[width=0.45\textwidth]{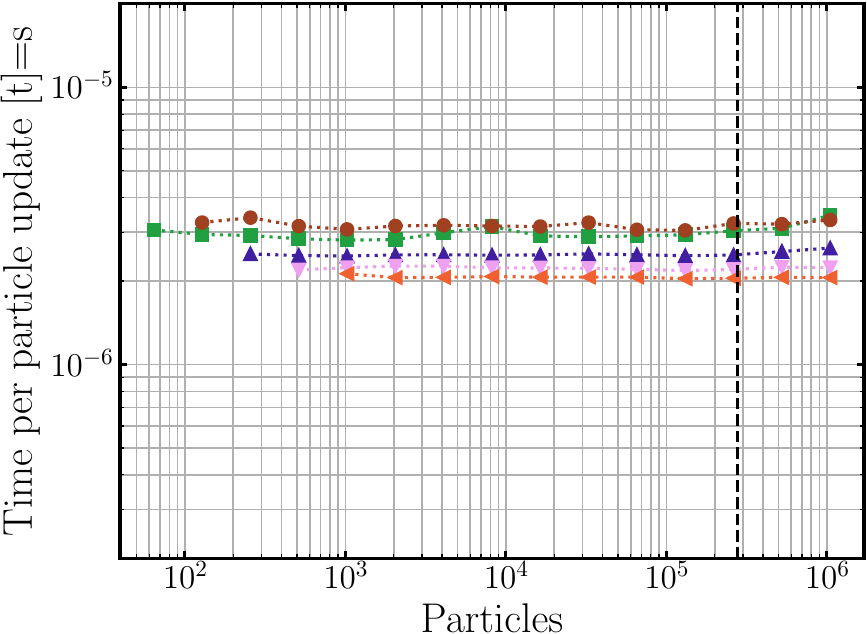}     
    \includegraphics[width=0.45\textwidth]{experiments/kernel-throughput-2/gn003/density_scattered_generic_no_dist_check.pdf}     
  \caption{
    \replaced[id=R2]{
      Comparison of time per particle update of the Hopper GPU (left) and Grace CPU (right) density kernel launches for various $ppc$.
      We use explicit mapping of data to the device, i.e.~copies, and the
      measurements incorporate all data movements, including data transfers to and from the accelerator. 
      We start from scattered data on the host and employ either the kernels with branching (top) or with masking (bottom).
      On the CPU, dashed lines are the vanilla version without compiler transformations, runtimes represented by dotted lines result from our compiler transformations.
      On the GPU, we exclusively rely on the data transformations.
   	}{
   	   	Cost per particle update for various kernels on Hopper GPU. 
   	All GPU kernels are generated through attribute annotations and transfer their data explicitly to the accelerator prior to the kernel start. 
   	In lexicographic order: $ppc=64$, $ppc=128$, $ppc=256$, and $ppc=1024$.
   }
   \label{figure:results:gpu}
  }
\end{figure}

\begin{figure}[htb]
\centering
    \includegraphics[width=0.6\textwidth]{experiments/kernel-throughput-2/gn003/cell_legend.pdf}
    \includegraphics[width=0.45\textwidth]{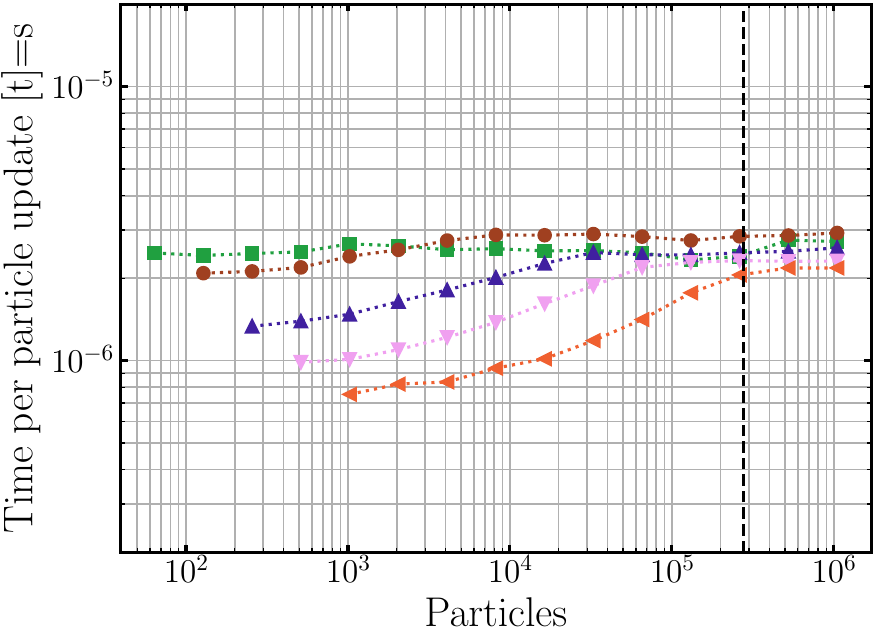}
    \includegraphics[width=0.45\textwidth]{experiments/kernel-throughput-2/gn003/density_gpu_scattered_generic_dist_check.pdf}     
    \includegraphics[width=0.45\textwidth]{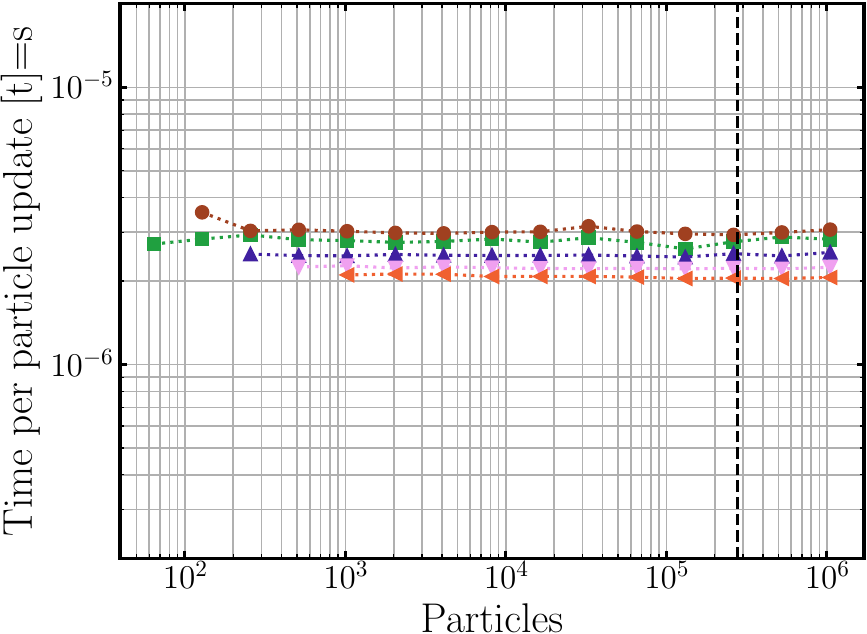}     
    \includegraphics[width=0.45\textwidth]{experiments/kernel-throughput-2/gn003/density_gpu_scattered_generic_no_dist_check.pdf}     
  \caption{
    \replaced[id=R2]{
      We repeat the measurements from Figure~\ref{figure:results:gpu} with explicit data mapping (old data right) to an implementation relying on the Grace Hopper's USM:
      The density kernel is benchmarked for a kernel employing branching (top) and a kernel masking out contributions (bottom). 
    }{
      Experiments from Figure~\ref{figure:results:gpu} repeated with shared memory, i.e.~without any explicit memory transfer.
    }
    \label{figure:results:gpu-shmem}
  }
\end{figure}

\begin{figure}[htb]
\centering
    \includegraphics[width=0.6\textwidth]{experiments/kernel-throughput-2/gn003/cell_legend.pdf}
    \includegraphics[width=0.45\textwidth]{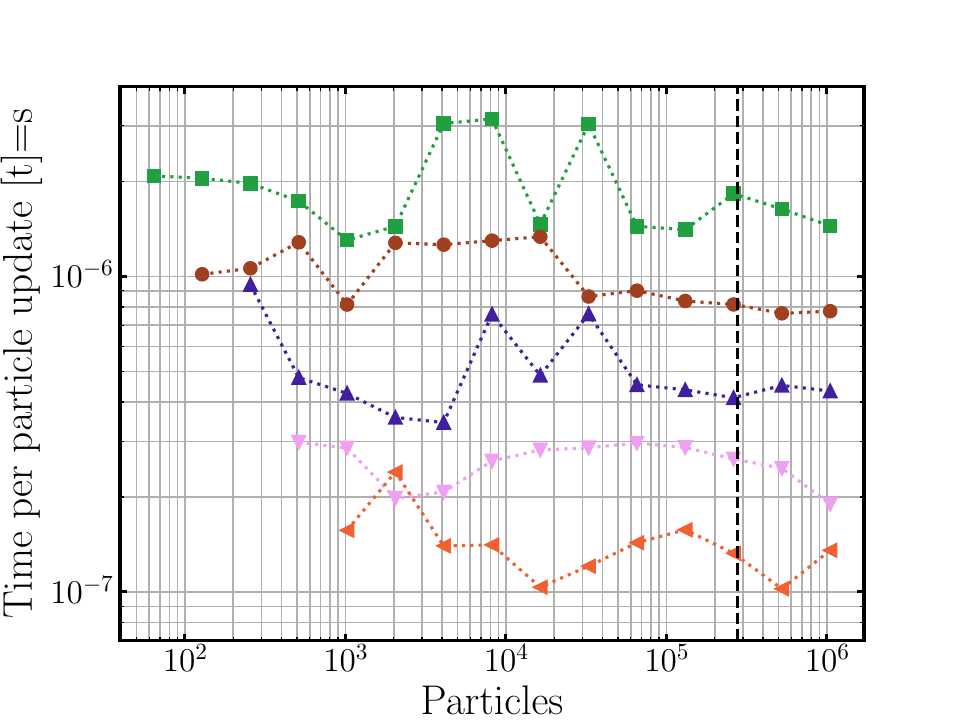}
    \includegraphics[width=0.45\textwidth]{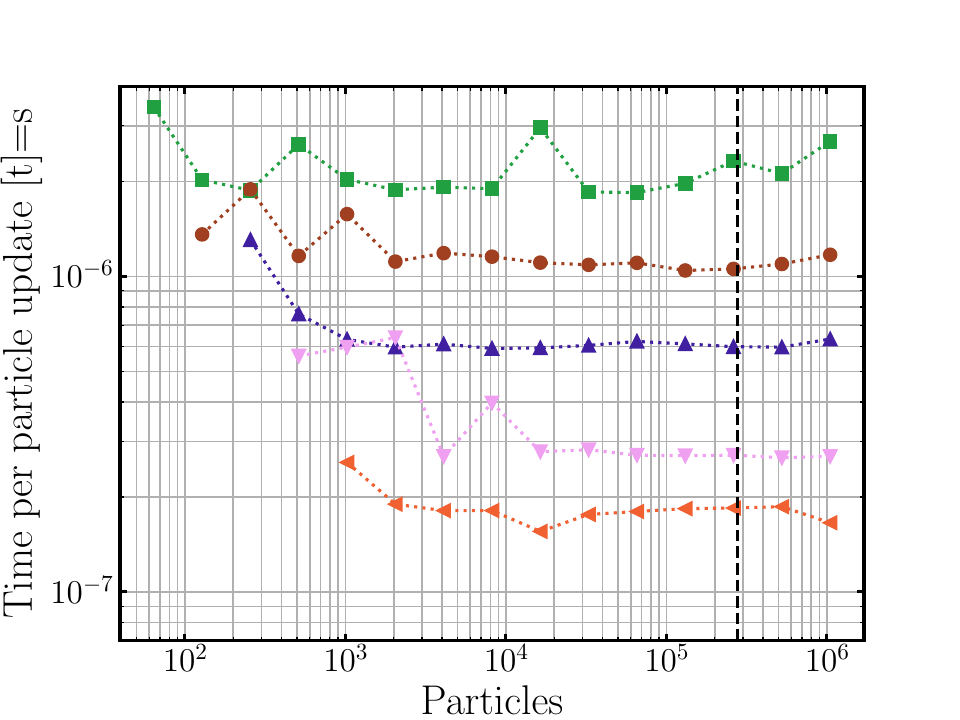}     
  \caption{
    \added[id=R2]{
     Unified shared memory (left) vs.~explicit data mapping (right) for the drift kernel on the Hopper GPU.
    }
   \label{figure:results:gpu-shmem-drift}
  }
\end{figure}

%
%
\added[id=Ours]{
 We benchmark the GPU data against the fastest CPU variant with scattered data.
 In return, we only benchmark kernels subject to the AoS-to-SoA transformations, since it is impractical to explicitly map scattered data onto the device.
}
The GPU throughput is low compared to the theoretical capability of the device, but the data is qualitatively similar to the results on the Intel and ARM CPUs (Figure~\ref{figure:results:gpu})\replaced[id=Ours]{: Larger total particle counts lead to a saturated cost per particle update, while small total counts make us benefit from local memory buffers. This effect however only arises for an implementation with branching}{ in that larger total particle counts give us a better throughput}.
We see a less pronounced relation of throughput to $ppc$, but there is an inversion of the previously stated trend:
\replaced[id=R1]{Larger $ppc$ counts yield a }{
 Bigger $ppc$ give us} superior time-to-solution.
The Grace Hopper's shared memory capabilities give us \replaced[id=R1]{negligible performance gain for the $\mathcal{O}(N^2)$ kernels but close to a speedup of two for the linear kernels}{another factor of two in performance} (Figure~\ref{figure:results:gpu-shmem}\added[id=R2]{ and Figure~\ref{figure:results:gpu-shmem-drift}}). 
\added[id=Ours]{
 For a sufficiently high number of particles and quadratic kernels, we observe that $ppc \geq 256$ perform better on the GPU than on the CPU regardless of the scheme used.
 The linear kernels suffer from GPU offloading.
}

%
%
GPUs require high concurrency levels in the compute kernels to unfold their full potential.
In return, they are not as vulnerable to the branching due to the distance checks as the vectorised CPU kernels\deleted[id=Ours]{:
The masking facilities on the accelerator can compensate for the thread divergence}.
Therefore, picking larger $ppc$ becomes reasonable.
\added[id=Ours]{
 Conversely, offloading helps if and only if the number of particles per cell is big.
}

Despite the tight memory connection on the Grace-Hopper superchip, we see a limited performance improvement in our kernels\added[id=Ours]{ due to GPU offloading overall}, which is due to the low concurrency.
It is not clear from the present setups, to which degree our data suffers from a lack of accelerator-specific tuning, such as the tailoring of warp sizes or simply too many, too small kernels.
Unless the GPU software stack or the hardware start to facilitate very small kernels, a direct mapping of compute kernels onto GPU kernels does not deliver satisfying results.

Our vanilla offloading in the compiler issues a sequence of three tasks: map, compute, map back.
They are connected through dependencies yet launched asynchronously.
After the final map, we wait.
If we enable the superchip's shared memory, the three tasks degenerate to one single blocking task.
Nevertheless, it seems that the chip delivers better performance in this mode.
We assume that OpenMP struggles to overlap the data transfer of all CPU threads firing map-compute-map sequences simultaneously.
Contrary, the hardware seems to succeed to let shared memory tasks run ahead and to bring in all required data lazily upon demand\added[id=Ours]{, which is an effect that is pronounced for linear kernels only}.

All data conversion resides on the CPU.
We may therefore conclude that any improvement of compute speed on the accelerator is (partially) consumed by data transfers or on GPU \replaced[id=Ours]{kernel dispatch}{} overheads.
It is also not clear what role the inferior support for FP64 on the GPU plays for \replaced[id=R2]{our}{out} experiments.
It is reasonable to believe that the insight might change qualitatively if we were able to convert our kernels to FP32---an endeavour which requires care as it tends to introduce numerical long-term instabilities.

\begin{observation}
 Our annotations streamline GPU offloading as they tackle the memory bottleneck, but they do not free the developer from generating tailored compute kernels with a sufficiently high concurrency level. 
\end{observation}

\noindent
It is well-know that GPUs benefit from gathering multiple compute tasks into one larger GPU kernel \cite{Nasar:2024:GPU}.
Similar techniques have been proposed in a more general way for task-based systems~\cite{LiShultzWeinzierl:2022:Tasking}.
Our data suggest that using such techniques \replaced[id=R1]{is}{are} absolutely essential to write fast GPU code, and that the naive offloading of individual, tiny compute kernels does not yield satisfying performance.
However, the core idea of reordering data through compiler annotations streamlines the writing of more complex (meta-)kernels as well.

\subsection{\added[id=R3]{Contextualisation and related work}}

%
%
\added[id=R3]{
 Discussions around \emph{temporary and local} AoS-to-SoA conversions typically focus on their suitability for particular hardware topologies \cite{Che:2011:Dymaxion,Gustavson:2012:InPlaceStorage,Hirzel:2007:DataLayoutForOO,Homann:2018:SoAx,Springer:2018:SoALayout} and emphasise that GPUs and CPUs require different storage schemes \cite{Che:2011:Dymaxion,Majeti:2014:CompilerDrivenDataLayoutTransformations,Strzodka:2011:AbstractionSoA,Sung:2012:DataLayoutTransformations}. 
 Our work extends beyond this approach by explicitly enabling format switching based on the program phase: 
 We propose that each kernel might operate on its own optimal data format, i.e.~we move away from a mindset where the optimal data format is statically bound to a specific device and data combination. 
 In this regard, our approach aligns with the philosophy behind Kokkos \cite{Trott:2022:Kokkos}, which maintains separation between execution spaces, data structures, and algorithms.
}


%
%
\added[id=R3]{
We do not yet optimise the conversion process, even though our data demonstrates that its overhead can be large. 
In the context of CPU-GPU data transfer, it is natural to consider fusing the conversion with the data transfer itself. 
On one hand, it may be reasonable to deploy the data reordering on the GPU, i.e.~to permute the operators in \eqref{equation:formalism:permute-cpu-gpu-conversion}. 
On the other hand, there is potential to overlap such conversion with the actual data copying by breaking down the data streams into chunks and running conversion routines in parallel \cite{Che:2011:Dymaxion}. 
Conversion and data movement operations inherently incorporate concurrency \cite{Gustavson:2012:InPlaceStorage}. 
Consequently, there is significant room for runtime improvement in our approach.
}

%
%
\added[id=R3]{
 We may assume that the best-performing codes select the optimal data format from the outset---one from which they deviate only during specific phases. 
  Identifying such an advantageous baseline data format is likely highly application-specific. 
  Our work challenges conventional wisdom for which type of kernels we benefit most from particular data layouts (cmp.~statements on streaming kernels~\cite{Intel:MemoryLayoutTransformations}), but does not yet provide any cost model for determining when switching data layouts is beneficial. 
  Insights on heuristics as well as critical analyses have been published, though they have not been incorporated into our solution \cite{Hundt:2006:StructureLayoutOptimisation,Majeti:2014:CompilerDrivenDataLayoutTransformations,Vikram:2014:LLVM,Xu:2014:SemiAutomaticComposition}. 
  Future implementations may need to optimise over both a \emph{static, global} optimisation of the baseline data arrangement and temporary, local conversions along a global cost model.
}

%
%
\added[id=R3]{
Our work is based on the premise that developers should be free to experiment with data formats, while the compiler can automate tasks such as identifying \emph{views}, handling the actual transformations, and creating bespoke, tailored compute kernels. 
An alternative to our \emph{guided transformations} deploying work to the compiler would be code written to be storage format-agnostic from the outset \cite{Childers:2024:CPPReflections,Homann:2018:SoAx,Jubertie:2018:DataLayoutAbstractionLayers,Springer:2018:SoALayout,Xu:2014:SemiAutomaticComposition}.
Choosing either approach is a matter of taste, but implementing code transformations exclusively within the compiler avoids situations where compilers fail to ``see through'' and optimise complex data access abstractions \cite{Jubertie:2018:DataLayoutAbstractionLayers}. 
It however remains an open question to what extent our techniques could be more tightly integrated with other compiler optimisation passes.
}

\section{Conclusion}
\label{section:conclusion}

\added[id=copy]{
 SoA is not always superior to AoS. 
 Tasks such as boundary data exchange in domain decomposition codes, particle movements over meshes, or, in
 general, any algorithm that has to alter or permute struct sequences or
 access it in a non-continuous way \cite{Strzodka:2011:AbstractionSoA} benefit
 from AoS.
 The size of characteristic sequences \cite{Homann:2018:SoAx} and the memory
 footprint per struct further affect which storage format performs better.
 Finally, any statement on a storage format superiority depends upon how successful the compiler 
 vectorises an algorithm
 \cite{Jubertie:2018:DataLayoutAbstractionLayers,Xu:2014:SemiAutomaticComposition} and what the target architecture looks like.
 The choice of an optimal data structure is context-dependent.
 There is no ``one format rules them all''.
} 
Our approach accommodates this insight as it strictly separates the optimisation over data layouts from the modelling.
By adding, modifying or removing annotations, users can investigate better-suited data layouts\added[id=Ours]{ and change them upon demand}.

Our experiments with temporary out-of-place transformations~\cite{Sung:2012:DataLayoutTransformations} and struct peeling~\cite{Hundt:2006:StructureLayoutOptimisation} suggest that annotations bringing data into the right format at the right time has significant performance optimisation potential.
However, our observations also suggest that they do not free developers from manual performance engineering.
Instead, they alter the character of such engineering.
Streamlining all data conversion allows the engineer to focus more on other aspects such as kernel-level vectorisation or the creation of GPU kernels handling larger data sets in one go without small kernel launches.

The observations also challenge the common knowledge that it is important to organise data carefully within large-scale simulation codes.
Instead, there is the potential to work with wildly scattered heap data and to consolidate such data just before the compute-heavy kernels are launched.
It is likely that this observation is not upheld for other codes and can be challenged.
However, it seems to hold for SPH.

Future work will have to push the notion of on-demand data transformations.
At the moment, our view constructions are strictly temporal and tied to individual loops.
If a force calculation follows a density compute step, we convert data forth and back, which is not necessary: 
We potentially could hold it in SoA, albeit with bigger $\mathbb{A}$ sets and likely not on the call stack anymore.
It will remain subject of future work to study how such considerations can feed into compiler-guided format conversions.
It is also important to continue to explore how annotations can help to \replaced[id=R1]{performance}{performanc} engineer modern codes efficiently, what role data conversions can play in this context, and how different aspects such as the conversion and multicore parallelisation or GPU offloading can be brought together.
As C starts to support annotations, too, such \replaced[id=Ours]{conversions}{conversations} generalise beyond the realms of C++ only.

\section*{Acknowledgments}

Tobias' research has been supported by EPSRC's Excalibur
programme through the DDWG projects \textit{PAX--HPC} (Gant EP/W026775/1)
as well as \textit{An ExCALIBUR Multigrid Solver Toolbox for ExaHyPE}
(EP/X019497/1).
His group appreciates the support by Intel's Academic Centre of
Excellence at Durham University. 
The work has received funding through the eCSE project ARCHER2-eCSE11-2
\emph{ExaHyPE-DSL}, and Pawel and Tobias have been supported through EPSRC's \emph{HAI-End} project.

The code development relied on experimental test nodes installed within  
the DiRAC@Durham facility managed by the Institute for Computational Cosmology
on behalf of the STFC DiRAC HPC Facility
(\href{www.dirac.ac.uk}{www.dirac.ac.uk}). The equipment was funded by BEIS capital funding via STFC capital grants ST/K00042X/1, ST/P002293/1, ST/R002371/1 and ST/S002502/1, Durham University and STFC operations grant ST/R000832/1. DiRAC is part of the National e-Infrastructure.

\bibliographystyle{splncs04}
\bibliography{paper}

\newpage

\appendix

\section{Compiler download}
\label{appendix:compiler-download}

The forked Clang/LLVM project is publicly available at
\linebreak 
\url{https://github.com/pradt2/llvm-project}, our extensions are available in the \texttt{hpc-ext} branch.

The \texttt{hpc-ext} branch is up to date with the upstream LLVM \texttt{main} branch as of 12 Feb 2025.

To avoid conflicts, it is strongly recommended to remove any existing Clang/LLVM installations before proceeding.

To build and install the compiler toolchain, clone or otherwise download and unpack the repository, create a build folder in the top-level repository directory, and execute all steps from Algorithm~\ref{algorithm:appendix:compiler-download:build-compiler}.

\begin{algorithm}[htb]
    \begin{algorithmic}[1]
      \State 
      \begin{verbatim}
cmake 
  -DCMAKE_BUILD_TYPE="RelWithDebInfo" 
  -DCMAKE_C_COMPILER="gcc" 
  -DCMAKE_CXX_COMPILER="g++" 
  -DLLVM_ENABLE_PROJECTS="clang;lld" 
  -DLLVM_ENABLE_RUNTIMES="openmp;offload" 
  -DLLVM_TARGETS_TO_BUILD="X86;AArch64;NVPTX;AMDGPU"
  ../llvm
      \end{verbatim}
\State \verb|make && make install|
    \end{algorithmic}
  \caption{
    Building and installing the compiler
    \label{algorithm:appendix:compiler-download:build-compiler}
    }
\end{algorithm}

\noindent

From hereon, \texttt{clang++} directs to our modified compiler version.


The command-line interface (CLI) is backwards compatible with the upstream
Clang/LLVM version. 
The support for the new attributes discussed in this paper is enabled by default, no additional compiler flags are needed.

If the use of any of the new attributes leads to a compilation error, a common
troubleshooting starting point is to inspect the rewritten source code. To see
the rewritten code, we can add \texttt{-fpostprocessing-output-dump} to the
compilation flags. This flag causes the post-processed source code be written to the standard output.

\section{Download, build and run testbench}
\label{appendix:download-and-compilation}

All of our code is hosted in a public git repository 
at \url{https://gitlab.lrz.de/hpcsoftware/Peano}. 
Our benchmark scripts are merged into the repository's
main, i.e.~all results can be reproduced with the main branch. 
Yet, to use the exact same code version as used for this paper, please switch to 
the \linebreak \texttt{particles-noh2D-vectorisation-2} branch.

\begin{algorithm}[htb]
    \begin{algorithmic}[1]
      \State \verb|git clone https://gitlab.lrz.de/hpcsoftware/Peano|
      \State \verb|cd Peano|
      \State \verb|libtoolize; aclocal; autoconf; autoheader|
      \State \verb|cp src/config.h.in .; automake --add-missing|
      \State \begin{verbatim}
./configure CXX=... CC=... CXXFLAGS=... LDFLAGS=... 
   --enable-loadbalancing 
   --enable-particles 
   --enable-swift 
   --with-multithreading=omp
\end{verbatim}
      \State \verb|make|
    \end{algorithmic}
  \caption{
    Cloning the repository and setting up the autotools environment. 
    \label{algorithm:appendix:download-and-compile:autotools}
    }
\end{algorithm}

The test benchmarks in the present paper are shipped as part of Peano's
benchmark suite, which implies that Peano has to be configured and built first.
The code base provides CMake and autotools (Alg.~\ref{algorithm:appendix:download-and-compile:autotools})
bindings.
Depending on the target platform, different compiler options have to be used.
Once configured, the build (\texttt{make}) yields a set of static libraries
providing the back-end of our benchmarks.

The actual benchmark can be found in Peano's subdirectory \linebreak
\texttt{benchmarks/swift2/hydro/kernel-throughput}.
Within this directory, call \linebreak \texttt{./benchmark.sh} to run the CPU-only benchamrks, and \texttt{./benchmark-offload.sh} to run the accelerator offloading dependent bechmarks.

Executing any of the benchmarking scripts produces a series of log files in the working directory are the raw data plotted and presented in this paper.

To reproduce the plots presented in this paper, call \linebreak \texttt{python3 ./print-kernel-throughput.py *.log} and \linebreak \texttt{python3 ./print-kernel-speedups.py ./} in the working directory where the log files are.

\end{document}